\documentclass[prd,aps,floatfix,superscriptaddress,twocolumn]{revtex4-2}
\usepackage{amsfonts,amsmath,amssymb,amsthm}
\usepackage{bm}
\usepackage{booktabs}
\usepackage{braket} % Dirac bra-ket
\usepackage{cases}
\usepackage{color}
\usepackage{dcolumn}
\usepackage{epsfig}
\usepackage{epstopdf}   %{feynmp}
\usepackage{graphicx}
\usepackage[colorlinks,citecolor=blue,anchorcolor=red,menucolor=red,linkcolor=red,filecolor=red,runcolor=red,urlcolor=blue,frenchlinks=red]{hyperref}
\usepackage{indentfirst}
\usepackage{latexsym}
\usepackage{longtable}
\usepackage{multirow}
\usepackage{rotating}
\usepackage{slashed}  %for Feynman symbols $\slashed{x}$
\usepackage{subfigure}
\newcommand{\etal}{\textit{et~al}.~}
\allowdisplaybreaks[4]
\begin{document}
%%%%%%%%%%%%%%%%%%%%%%%%%%%%%%%%%%%%%%%%%%%%%%%%%%%%%%%%%%%%%%%%%%%%
\title{Doubly heavy tetraquarks in an extended chromomagnetic model}
%%%%%%%%%%%%%%%%%%%%%%%%%%%%%%%%%%%%%%%%%%%%%%%%%%%%%%%%%%%%%%%%%%%%
%\author{Xin-Zhen Weng (翁新震)}
\author{Xin-Zhen Weng}
\email{xzhweng@pku.edu.cn}
\affiliation{Center of High Energy Physics, Peking University, Beijing 100871, China}
\affiliation{School of Physics and Astronomy, Tel Aviv University, Tel Aviv 69978, Israel}
%%%
%\author{Wei-Zhen Deng (邓卫真)}
\author{Wei-Zhen Deng}
\email{dwz@pku.edu.cn}
\affiliation{School of Physics and State Key Laboratory of Nuclear Physics and Technology, Peking University, Beijing 100871, China}
%%%
%\author{Shi-Lin Zhu (朱世琳)}
\author{Shi-Lin Zhu}
\email{zhusl@pku.edu.cn}
\affiliation{Center of High Energy Physics, Peking University, Beijing 100871, China}
\affiliation{School of Physics and State Key Laboratory of Nuclear Physics and Technology, Peking University, Beijing 100871, China}
\affiliation{Collaborative Innovation Center of Quantum Matter, Beijing 100871, China}
%%%%%%%%%%%%%%%%%%%%%%%%%%%%%%%%%%%%%%%%%%%%%%%%%%%%%%%%%%%%%%%%%%%%
\date{\today}
%%%%%%%%%%%%%%%%%%%%%%%%%%%%%%%%%%%%%%%%%%%%%%%%%%%%%%%%%%%%%%%%%%%%
\begin{abstract}
%%%%%%%%%%%%%%%%%%%%%%%%%%%%%%%%%%%%%%%%%%%%%%%%%%%%%%%%%%%%%%%%%%%%

Using an extended chromomagnetic model, we perform a systematic study of the masses of the doubly heavy tetraquarks.
We find that the ground states of the doubly heavy tetraquarks are dominated by color-triplet $\ket{(qq)^{\bar{3}_{c}}(\bar{Q}\bar{Q})^{3_{c}}}$ configuration, which is opposite to that of the fully heavy tetraquarks.
The combined results suggest that the color-triplet configuration becomes more important when the mass difference between the quarks and antiquarks increases.
We find three stable states which lie below the thresholds of two pseudoscalar mesons.
They are the $IJ^{P}=01^{+}$ $nn\bar{b}\bar{b}$ tetraquark, the $IJ^{P}=00^{+}$ $nn\bar{c}\bar{b}$ tetraquark and the $J^{P}=1^{+}$ $ns\bar{b}\bar{b}$ tetraquark.
%

%%%%%%%%%%%%%%%%%%%%%%%%%%%%%%%%%%%%%%%%%%%%%%%%%%%%%%%%%%%%%%%%%%%%
\end{abstract}
%%%%%%%%%%%%%%%%%%%%%%%%%%%%%%%%%%%%%%%%%%%%%%%%%%%%%%%%%%%%%%%%%%%%

\maketitle
\thispagestyle{empty} % 首页不显示页码

%%%%%%%%%%%%%%%%%%%%%%%%%%%%%%%%%%%%%%%%%%%%%%%%%%%%%%%%%%%%%%%%%%%%
\section{Introduction}
\label{Sec:Introduction}
%%%%%%%%%%%%%%%%%%%%%%%%%%%%%%%%%%%%%%%%%%%%%%%%%%%%%%%%%%%%%%%%%%%%

Besides the conventional meson and baryon composed of quark-antiquark pair and three quarks, there also exist hadrons composed of more than three quarks, or gluons.
These states are called exotic states, such as the 
tetraquark~\cite{Cui:2006mp,Park:2013fda}, 
pentaquark~\cite{Wu:2010jy,Liu:2015fea}, 
molecule~\cite{Swanson:2003tb,Carames:2010zz,Chen:2015add}, 
glueball~\cite{Carlson:1982er,Chen:2021cjr}, 
hybrid~\cite{Zhu:2005hp,Esposito:2016itg}, etc.
In 2003, the first charmoniumlike state $X(3872)$ was observed by the Belle Collaboration in the exclusive $B^{\pm}{\rightarrow}K^{\pm}\pi^{+}\pi^{-}J/\psi$ decays~\cite{Choi:2003ue}.
Its quantum number is $I^{G}J^{PC}=0^{+}1^{++}$~\cite{Zyla:2020zbs}.
The discovery of $X(3872)$ opens a new era of the hadron spectroscopy.
Lots of charmoniumlike and bottomoniumlike states were found since then, such as the $Y(4260)$~\cite{Aubert:2005rm}, $Z_{c}(3900)$~\cite{Ablikim:2013mio,Liu:2013dau}, $Z_{b}(10610)$ and $Z_{b}(10650)$~\cite{Bondar:2011aa} states.
In the fully heavy sector, the LHCb collaboration observed a narrow structure and a wide structure in the $J/\psi$-pair invariant mass spectrum in the range of $6.2\sim7.2~\text{GeV}$, which could be all-charm hadrons~\cite{Aaij:2020fnh}.
More details can be found in Refs.~\cite{Chen:2016qju,Esposito:2016noz,Lebed:2016hpi,Ali:2017jda,Guo:2017jvc,Yuan:2018inv,Brambilla:2019esw,Liu:2019zoy} and references therein.

The heavy quarks in these states are all in hidden flavor(s).
%
%There also exist exotic states with open heavy flavor(s).
%
In 2017, the LHCb Collaboration observed the $\Xi_{cc}^{++}$ in the $\Lambda_c^+K^-\pi^+\pi^+$ decay channel~\cite{Aaij:2017ueg}.
Its mass was determined to be $3621.40\pm0.72(\text{stat.})
\pm0.27(\text{syst.}) \pm0.14(\Lambda_c^+)~\text{MeV}$.
This is the first doubly heavy baryon observed in experiment.
The $\Xi_{cc}^{++}$ baryon gives implications for the doubly heavy tetraquarks~\cite{Eichten:2017ffp,Karliner:2017qjm}, which are exotic states with open heavy flavors.
Recently, the LHCb Collaboration observed a very narrow state in the $D^{0}D^{0}\pi^{+}$ mass spectrum~\cite{LHCb:2021vvq,LHCb:2021auc,Muheim-2021-HighlightsfromtheLHCbExperiment,Polyakov-2021-RecentLHCbResultsonExoticMesonCandidates}.
Under the $J^{P}=1^{+}$ assumption, its mass with respect to the $D^{*+}D^{0}$ and width are
\begin{equation}
	\delta{m}_{\text{BW}}
	=
	-273\pm61\pm5_{-14}^{+11}~\text{keV}\,,
\end{equation}
and
\begin{equation}
	\Gamma_{\text{BW}}
	=
	410\pm165\pm43_{-38}^{+18}~\text{keV}\,,
\end{equation}
respectively.
The statistic significance for the signal is over $10\sigma$, while that for $\delta{m}_{\text{BW}}<0$ is $4.3\sigma$.
This structure is consistent with the $DD^{*}$ molecule interpretation predicted by Li~\etal within the one-boson-exchange (OBE) model~\cite{Li:2012ss}.
The discovery of the $T_{cc}^{+}$ inspired many related studies~\cite{Agaev:2021vur,Chen:2021vhg,Dai:2021wxi,Dong:2021bvy,Feijoo:2021ppq,Ling:2021bir,Li:2021zbw,Meng:2021jnw,Wu:2021kbu,Yan:2021wdl}.
Actually, the doubly heavy tetraquarks have been studied extensively in the literature, with models like 
the quark model~\cite{Ader:1981db,Carlson:1987hh,
Zouzou:1986qh,
Silvestre-Brac:1993zem,
Gelman:2002wf,
Vijande:2003ki,
Janc:2004qn,
Cui:2006mp,Vijande:2006jf,
Vijande:2007ix,Ebert:2007rn,
Lee:2009rt,Yang:2009zzp,
Valcarce:2010zs,
Luo:2017eub,
Deng:2018kly,Park:2018wjk,Yan:2018gik,
Maiani:2019lpu,Yang:2019itm,Zhu:2019iwm,
Braaten:2020nwp,Cheng:2020wxa,Lu:2020rog,Meng:2020knc,Tan:2020ldi,
Faustov:2021hjs,Meng:2021yjr,Noh:2021lqs},
QCD sum rules~\cite{Navarra:2007yw,
	Wang:2010uf,
	Dias:2011mi,
	Du:2012wp,
	Chen:2013aba,
	Wang:2017dtg,Wang:2017uld,
	Agaev:2019kkz,Agaev:2019qqn,Tang:2019nwv,
	Wang:2020jgb},
Lattice QCD~\cite{Wagner:2011ev,
Bicudo:2012qt,
Brown:2012tm,
Ikeda:2013vwa,
Guerrieri:2014nxa,
Bicudo:2015kna,Bicudo:2015vta,
Peters:2016isf,Bicudo:2016jwl,Bicudo:2016ooe,Francis:2016hui,Francis:2016nmj,
Bicudo:2017szl,Cheung:2017tnt,
Francis:2018jyb,Junnarkar:2018twb,
Leskovec:2019ioa,
Hudspith:2020tdf,Mohanta:2020eed},
the OBE potentials~\cite{Ericson:1993wy,Tornqvist:1993ng,
Ding:2009vj,
Ohkoda:2012hv,
Li:2012ss,
Ding:2021igr},
chiral perturbation theory~\cite{Manohar:1992nd,
Liu:2012vd,
Xu:2017tsr,
Wang:2018atz},
etc.
Their production has also been studied (for instance, see Refs.~\cite{Esposito:2013fma,Qin:2020zlg}).
The studies suggest that the masses of some of the doubly heavy tetraquarks are lighter than the thresholds of two mesons, which makes them stable against the strong and electromagnetic decays.
For example, Du~\etal~\cite{Du:2012wp} studied the $QQ\bar{q}\bar{q}'$ ($Q=c,b$ and $q,q'=u,d,s$) in the QCD sum rules.
They found that the $bb\bar{q}\bar{q}'$'s are stable.
The stableness of doubly heavy tetraquarks is also supported by the lattice QCD calculations.
Leskovec~\etal~\cite{Leskovec:2019ioa} used lattice QCD to investigate the spectrum of the $\bar{b}\bar{b}ud$ four-quark system with quantum numbers $I(J^{P})=0(1^+)$, and obtained a binding energy of $(-128\pm24\pm10)~\text{MeV}$, corresponding to the mass $10476\pm24\pm10~\text{MeV}$.
Mohanta and Basak~\cite{Mohanta:2020eed} studied the $bb\bar{u}\bar{d}$ states on lattice using non relativistic QCD (NRQCD) action for bottom and highly improved staggered quark (HISQ) action for the light up/down quarks.
They got the binding energy for the $1^{+}$ $bb\bar{u}\bar{d}$ tetraquark system to be
$-189(18)~\text{MeV}$ compared to the $BB^{*}$.
Using $\Xi_{cc}^{++}$ mass as input, Karliner and Rosner~\cite{Karliner:2017qjm} predicted that the mass of the ground state of $IJ^{P}=01^{+}$ doubly charm tetraquark ($T_{cc}$) to be $3882.2\pm12~\text{MeV}$ in the chromomagnetic model.

In the quark model~\cite{Eichten:1978tg,Isgur:1977ef,Godfrey:1985xj,Capstick:1986bm,DeRujula:1975qlm}, the mass of hadron can be decomposed into the quark masses, the kinetic energy and the potentials which include the color-independent Coulomb and confinement interactions, and the hyperfine interactions like the spin-spin, spin-orbit, and tensor terms.
If we restrict to the $S$-wave states, the spin-orbit and tensor interactions do not contribute.
We can use the extended chromomagnetic model~\cite{Sakharov:1966tua,DeRujula:1975qlm,Jaffe:1976ig,Jaffe:1976ih,Cui:2006mp,Buccella:2006fn,Hogaasen:2013nca,Karliner:2014gca,Luo:2017eub,Weng:2018mmf,Weng:2019ynv,Weng:2020jao}.
In this model, the masses of $S$-wave hadrons consist of effective quark masses, the color interaction and the chromomagnetic interaction.
This simplified model gives good account of all $S$-wave mesons and baryons~\cite{Weng:2018mmf}.
In this work, we use the extended chromomagnetic model to study the $S$-wave doubly heavy tetraquarks.
With the wave function obtained, we further use a simple method to estimate the partial decay ratios of the tetraquark states.
In Sec.~\ref{Sec:Model} we introduce the methods of present work,
The numerical results are presented and discussed in
Sec.~\ref{Sec:Result}.
We conclude in Sec.~\ref{Sec:Conclusion}.
%

%%%%%%%%%%%%%%%%%%%%%%%%%%%%%%%%%%%%%%%%%%%%%%%%%%%%%%%%%%%%%%%%%%%%
\section{Formalism}
\label{Sec:Model}
%%%%%%%%%%%%%%%%%%%%%%%%%%%%%%%%%%%%%%%%%%%%%%%%%%%%%%%%%%%%%%%%%%%%

%%%%%%%%%%%%%%%%%%%%%%%%%%%%%%%%%%%%%%%%%%%%%%%%%%%%%%%%%%%%%%%%%%%%
\subsection{Hamiltonian}
\label{sec:Hamiltonian}
%%%%%%%%%%%%%%%%%%%%%%%%%%%%%%%%%%%%%%%%%%%%%%%%%%%%%%%%%%%%%%%%%%%%

%%%%%%%%%%%%%%%%%%%%%%%%%%%%%%%%%%%%%%%%%%%%%%%%%%%%%%%%%%%%%%%%%%%%
%\subsection{The Hamiltonian}
%\label{sec:hamiltonian}
%%%%%%%%%%%%%%%%%%%%%%%%%%%%%%%%%%%%%%%%%%%%%%%%%%%%%%%%%%%%%%%%%%%%

In the chromomagnetic model, the Hamiltonian of the $S$-wave hadron reads~\cite{Chan:1978nk,Fukugita:1978sn,Chao:1979mm,Chao:1980dv,Hogaasen:2013nca,Weng:2018mmf,Weng:2019ynv,Weng:2020jao}
\begin{equation}\label{eqn:ECM}
H
%%%
=
\sum_{i}m_{i}+H_{\text{CE}}+H_{\text{CM}}
\end{equation}
where $m_i$ is the effective mass of $i$th quark, $H_{\text{CE}}$ is the chromoelectric (CE) interaction~\cite{Hogaasen:2013nca,Weng:2018mmf,Weng:2019ynv,Weng:2020jao}
\begin{equation}\label{eqn:ECM:CE}
H_{\text{CE}}
%%%
=
-
\sum_{i<j}
a_{ij}
\bm{F}_{i}\cdot\bm{F}_{j}\,,
\end{equation}
and $H_{\text{CM}}$ is the chromomagnetic (CM) interaction~\cite{Jaffe:1976ig,Jaffe:1976ih,Cui:2006mp,Buccella:2006fn,Liu:2019zoy}
\begin{equation}\label{eqn:ECM:CM}
H_{\text{CM}}
%%%
=
-
\sum_{i<j}
v_{ij}
\bm{S}_{i}\cdot\bm{S}_{j}
\bm{F}_{i}\cdot\bm{F}_{j}\,.
\end{equation}
Here, $a_{ij}$ and $v_{ij}\propto\braket{\alpha_{s}(r_{ij})\delta(\mathbf{r}_{ij})}/m_{i}m_{j}$ are effective coupling constants which depend on the constituent quark masses and the spatial wave function.
$\bm{S}_{i}=\bm{\sigma}_i/2$ and $\bm{F}_{i}={\bm{\lambda}}_i/2$ are the quark spin and  color operators.
For the antiquark,
\begin{equation}
\bm{S}_{\bar{q}}=-\bm{S}_{q}^{*}\,,
\quad
\bm{F}_{\bar{q}}=-\bm{F}_{q}^{*}\,.
\end{equation}

Since
\begin{align}\label{eqn:m+color=color}
&
\sum_{i<j}
\left(m_i+m_j\right)
\bm{F}_{i}\cdot\bm{F}_{j}
\notag\\
%%%
={}&
\left(\sum_{i}m_{i}\bm{F}_i\right)
\cdot
\left(\sum_{i}\bm{F}_{i}\right)
-
\frac{4}{3}
\sum_{i}
m_{i}\,,
\end{align}
and the total color operator $\sum_i\bm{F}_i$ nullifies any color-singlet physical state, we can rewrite the Hamiltonian as~\cite{Weng:2018mmf,Weng:2019ynv,Weng:2020jao}
\begin{equation}\label{eqn:hamiltonian:final}
H=
-\frac{3}{4}
\sum_{i<j}m_{ij}V^{\text{C}}_{ij}
-
\sum_{i<j}v_{ij}V^{\text{CM}}_{ij} \,,
\end{equation}
where
\begin{equation}\label{eqn:para:color+m}
m_{ij}
=
\left(m_i+m_j\right)
+
\frac{4}{3}
a_{ij}\,,
\end{equation}
is the quark pair mass parameter.
$V^{\text{C}}_{ij}=\bm{F}_{i}\cdot\bm{F}_{j}$ and $V^{\text{CM}}_{ij}=\bm{S}_{i}\cdot\bm{S}_{j}\bm{F}_{i}\cdot\bm{F}_{j}$ are the color and CM interactions between quarks.
%

%%%%%%%%%%%%%%%%%%%%%%%%%%%%%%%%%%%%%%%%%%%%%%%%%%%%%%%%%%%%%%%%%%%%
\subsection{Wave function}
\label{sec:wavefunc}
%%%%%%%%%%%%%%%%%%%%%%%%%%%%%%%%%%%%%%%%%%%%%%%%%%%%%%%%%%%%%%%%%%%%

To investigate the masses of the tetraquarks, we need to construct the wave functions.
The total wave function is a direct product of the orbital, color, spin and flavor wave 
functions.
Here, the orbital wave function is symmetric since we only consider the $S$-wave states.
Since the Hamiltonian does not contain a flavor operator explicitly, we first construct the color-spin wave function, and then incorporate the flavor 
wave function to account for the Pauli principle.

The spins of the tetraquarks can be $0$, $1$ and $2$.
In the $qq{\otimes}\bar{q}\bar{q}$ configuration, the possible color-spin wave functions 
$\{\alpha_{i}^{J}\}$ are listed as follows,
\begin{enumerate}
\item $J^{P}=0^{+}$:
\begin{align}
&\alpha_{1}^{0}=\ket{\left(q_1q_2\right)_{1}^{6}\left(\bar{q}_{3}\bar{q}_{4}\right)_{1}^{\bar{6}}}_{0},
\notag\\
&\alpha_{2}^{0}=\ket{\left(q_1q_2\right)_{0}^{6}\left(\bar{q}_{3}\bar{q}_{4}\right)_{0}^{\bar{6}}}_{0},
\notag\\
&\alpha_{3}^{0}=\ket{\left(q_1q_2\right)_{1}^{\bar{3}}\left(\bar{q}_{3}\bar{q}_{4}\right)_{1}^{3}}_{0},
\notag\\
&\alpha_{4}^{0}=\ket{\left(q_1q_2\right)_{0}^{\bar{3}}\left(\bar{q}_{3}\bar{q}_{4}\right)_{0}^{3}}_{0},
\end{align}
\item $J^{P}=1^{+}$:
\begin{align}
&\alpha_{1}^{1}=\ket{\left(q_1q_2\right)_{1}^{6}\left(\bar{q}_{3}\bar{q}_{4}\right)_{1}^{\bar{6}}}_{1},
\notag\\
&\alpha_{2}^{1}=\ket{\left(q_1q_2\right)_{1}^{6}\left(\bar{q}_{3}\bar{q}_{4}\right)_{0}^{\bar{6}}}_{1},
\notag\\
&\alpha_{3}^{1}=\ket{\left(q_1q_2\right)_{0}^{6}\left(\bar{q}_{3}\bar{q}_{4}\right)_{1}^{\bar{6}}}_{1},
\notag\\
&\alpha_{4}^{1}=\ket{\left(q_1q_2\right)_{1}^{\bar{3}}\left(\bar{q}_{3}\bar{q}_{4}\right)_{1}^{3}}_{1},
\notag\\
&\alpha_{5}^{1}=\ket{\left(q_1q_2\right)_{1}^{\bar{3}}\left(\bar{q}_{3}\bar{q}_{4}\right)_{0}^{3}}_{1},
\notag\\
&\alpha_{6}^{1}=\ket{\left(q_1q_2\right)_{0}^{\bar{3}}\left(\bar{q}_{3}\bar{q}_{4}\right)_{1}^{3}}_{1},
\end{align}
\item $J^{P}=2^{+}$:
\begin{align}
&\alpha_{1}^{2}=\ket{\left(q_1q_2\right)_{1}^{6}\left(\bar{q}_{3}\bar{q}_{4}\right)_{1}^{\bar{6}}}_{2},
\notag\\
&\alpha_{2}^{2}=\ket{\left(q_1q_2\right)_{1}^{\bar{3}}\left(\bar{q}_{3}\bar{q}_{4}\right)_{1}^{3}}_{2},
\end{align}
\end{enumerate}
where the superscript $3$, $\bar{3}$, $6$ or $\bar{6}$ denotes the color, and the subscript 
$0$, $1$ or $2$ denotes the spin.

Next we consider the flavor wave function.
There are six types of total wave functions when we consider the Pauli principle:
\begin{enumerate}
\item Type A: $\varphi_{\text{A}}=\{(nn\bar{Q}\bar{Q})^{I=1},ss\bar{Q}\bar{Q}\}$
\begin{enumerate}
\item $J^{P}=0^{+}$:
\begin{align}\label{eqn:wavefunc:total:A0}
\Psi_{\text{A1}}^{0^{+}}
=
\varphi_{\text{A}}\otimes\alpha^{0}_{2},
\notag\\
%%%
\Psi_{\text{A2}}^{0^{+}}
=
\varphi_{\text{A}}\otimes\alpha^{0}_{3},
\end{align}
\item $J^{P}=1^{+}$:
\begin{align}\label{eqn:wavefunc:total:A1}
\Psi_{\text{A}}^{1^{+}}
=
\varphi_{\text{A}}\otimes\alpha^{1}_{4},
\end{align}
\item $J^{P}=2^{+}$:
\begin{align}\label{eqn:wavefunc:total:A2}
\Psi_{\text{A}}^{2^{+}}
=
\varphi_{\text{A}}\otimes\alpha^{2}_{2},
\end{align}
\end{enumerate}
\item Type B: $\varphi_{\text{B}}=\{(nn\bar{Q}\bar{Q})^{I=0}\}$
\begin{enumerate}
\item $J^{P}=1^{+}$:
\begin{align}\label{eqn:wavefunc:total:B1}
\Psi_{\text{B1}}^{1^{+}}
=
\varphi_{\text{B}}\otimes\alpha^{1}_{2},
\notag\\
%%%
\Psi_{\text{B2}}^{1^{+}}
=
\varphi_{\text{B}}\otimes\alpha^{1}_{6},
\end{align}
\end{enumerate}
\item Type C: $\varphi_{\text{C}}=\{(nn\bar{c}\bar{b})^{I=1},ss\bar{c}\bar{b}\}$
\begin{enumerate}
\item $J^{P}=0^{+}$:
\begin{align}\label{eqn:wavefunc:total:C0}
\Psi_{\text{C1}}^{0^{+}}
=
\varphi_{\text{C}}\otimes\alpha^{0}_{2},
\notag\\
%%%
\Psi_{\text{C2}}^{0^{+}}
=
\varphi_{\text{C}}\otimes\alpha^{0}_{3},
\end{align}
\item $J^{P}=1^{+}$:
\begin{align}\label{eqn:wavefunc:total:C1}
\Psi_{\text{C1}}^{1^{+}}
=
\varphi_{\text{C}}\otimes\alpha^{1}_{3},
\notag\\
%%%
\Psi_{\text{C2}}^{1^{+}}
=
\varphi_{\text{C}}\otimes\alpha^{1}_{4},
\notag\\
%%%
\Psi_{\text{C3}}^{1^{+}}
=
\varphi_{\text{C}}\otimes\alpha^{1}_{5},
\end{align}
\item $J^{P}=2^{+}$:
\begin{align}\label{eqn:wavefunc:total:C2}
\Psi_{\text{C}}^{2^{+}}
=
\varphi_{\text{C}}\otimes\alpha^{2}_{2},
\end{align}
\end{enumerate}
\item Type D: $\varphi_{\text{D}}=\{(nn\bar{c}\bar{b})^{I=0}\}$
\begin{enumerate}
\item $J^{P}=0^{+}$:
\begin{align}\label{eqn:wavefunc:total:D0}
\Psi_{\text{D1}}^{0^{+}}
=
\varphi_{\text{D}}\otimes\alpha^{0}_{1},
\notag\\
%%%
\Psi_{\text{D2}}^{0^{+}}
=
\varphi_{\text{D}}\otimes\alpha^{0}_{4},
\end{align}
\item $J^{P}=1^{+}$:
\begin{align}\label{eqn:wavefunc:total:D1}
\Psi_{\text{D1}}^{1^{+}}
=
\varphi_{\text{D}}\otimes\alpha^{1}_{1},
\notag\\
%%%
\Psi_{\text{D2}}^{1^{+}}
=
\varphi_{\text{D}}\otimes\alpha^{1}_{2},
\notag\\
%%%
\Psi_{\text{D3}}^{1^{+}}
=
\varphi_{\text{D}}\otimes\alpha^{1}_{6},
\end{align}
\item $J^{P}=2^{+}$:
\begin{align}\label{eqn:wavefunc:total:D2}
\Psi_{\text{D}}^{2^{+}}
=
\varphi_{\text{D}}\otimes\alpha^{2}_{1},
\end{align}
\end{enumerate}
\item Type E: $\varphi_{\text{E}}=\{ns\bar{Q}\bar{Q}\}$
\begin{enumerate}
\item $J^{P}=0^{+}$:
\begin{align}\label{eqn:wavefunc:total:E0}
\Psi_{\text{E1}}^{0^{+}}
=
\varphi_{\text{E}}\otimes\alpha^{0}_{2},
\notag\\
%%%
\Psi_{\text{E2}}^{0^{+}}
=
\varphi_{\text{E}}\otimes\alpha^{0}_{3},
\end{align}
\item $J^{P}=1^{+}$:
\begin{align}\label{eqn:wavefunc:total:E1}
\Psi_{\text{E1}}^{1^{+}}
=
\varphi_{\text{E}}\otimes\alpha^{1}_{2},
\notag\\
%%%
\Psi_{\text{E2}}^{1^{+}}
=
\varphi_{\text{E}}\otimes\alpha^{1}_{4},
\notag\\
%%%
\Psi_{\text{E3}}^{1^{+}}
=
\varphi_{\text{E}}\otimes\alpha^{1}_{6},
\end{align}
\item $J^{P}=2^{+}$:
\begin{align}\label{eqn:wavefunc:total:E2}
\Psi_{\text{E}}^{2^{+}}
=
\varphi_{\text{E}}\otimes\alpha^{2}_{2},
\end{align}
\end{enumerate}
\item Type F: $\varphi_{\text{F}}=\{ns\bar{c}\bar{b}\}$
\begin{align}\label{eqn:wavefunc:total:FJ}
\Psi_{\text{F}i}^{J^{+}}
=
\varphi_{\text{F}}\otimes\alpha^{J}_{i},
\end{align}
\end{enumerate}
Diagonalizing the Hamiltonian [Eq.~\eqref{eqn:hamiltonian:final}] in these bases, we can 
obtain the masses and eigenvectors of the doubly heavy tetraquarks.
%

%%%%%%%%%%%%%%%%%%%%%%%%%%%%%%%%%%%%%%%%%%%%%%%%%%%%%%%%%%%%%%%%%%%%
\subsection{Partial decay rates}
\label{sec:decay}
%%%%%%%%%%%%%%%%%%%%%%%%%%%%%%%%%%%%%%%%%%%%%%%%%%%%%%%%%%%%%%%%%%%%

Next we consider the strong decay properties of the tetraquarks.
There are various methods for studying the tetraquark decays, such as the dimeson decay through the quark interchange model~\cite{Liu:2014eka,Wang:2018pwi,Xiao:2019spy,Wang:2020prk} and the dibaryon decay through the $^{3}{P}_{0}$ model~\cite{Ader:1979yk,Barbour:1980vq,Roberts:1990ky,Liu:2016sip}.
These models require the dynamical structure of the hadrons, which is beyond the power of the chromomagnetic model.
Here we adopt a simple method to estimate the partial decay ratios of the tetraquark states.

In Sec.~\ref{sec:wavefunc} we have constructed the wave function in the $qq{\otimes}\bar{q}\bar{q}$ configuration, the tetraquark states are superposition of the bases.
The tetraquark states can also be written as the linear superposition of the bases in the $q\bar{q}{\otimes}q\bar{q}$ configuration (see Appendix~\ref{app:wavefunc:13x24}).
Normally, the $q\bar{q}$ component in the tetraquark can be either of color-singlet or of color-octet.
The former one can easily dissociate into two $S$-wave mesons in relative $S$ wave, which is called ``Okubo-Zweig-Iizuka- (OZI-)superallowed'' decays.
The recoupling coefficient tell us the overlap between the tetraquark and a particular meson~$\times$~meson state.
Then we can determine the decay amplitude of the tetraquark into that particular meson~$\times$~meson channel.
The latter one can only fall apart through the gluon exchange~\cite{Jaffe:1976ig,Strottman:1979qu}.
In this work, we will focus on the ``OZI-superallowed'' decays.

For each decay mode, the branching fraction is proportional to the square of the coefficient $c_i$ of the corresponding component in the eigenvectors, and also depends on the phase space.
For two body decay through $L$-wave, the partial decay width reads~\cite{Gao-1992-Group,Weng:2019ynv}
\begin{equation}\label{eqn:width}
	\Gamma_{i}=\gamma_{i}\alpha\frac{k^{2L+1}}{m^{2L}}{\cdot}|c_i|^2,
\end{equation}
where $m$ is the mass of the initial state, $k$ is the momentum of the final states in the rest frame of the initial state, $\alpha$ is an effective coupling constant, and $\gamma_{i}$ is a quantity determined by the decay dynamics.
Generally, $\gamma_{i}$ is determined by the spatial wave functions of both initial and final states, which are different for each decay process.
In the quark model, the spatial wave functions of the pseudoscalar and vector mesons are the same.
Thus for each tetraquark, we have
\begin{equation}
\gamma_{M_{1}M_{2}}
=
\gamma_{M_{1}M_{2}^{*}}
=
\gamma_{M_{1}^{*}M_{2}}
=
\gamma_{M_{1}^{*}M_{2}^{*}}
\end{equation}
where $M_{i}$ and $M_{i}^{*}$ are pseudoscalar and vector mesons respectively.
Then we can estimate the partial decay width ratios of the tetraquark states.
%

%%%%%%%%%%%%%%%%%%%%%%%%%%%%%%%%%%%%%%%%%%%%%%%%%%%%%%%%%%%%%%%%%%%%
\section{Numerical results}
\label{Sec:Result}
%%%%%%%%%%%%%%%%%%%%%%%%%%%%%%%%%%%%%%%%%%%%%%%%%%%%%%%%%%%%%%%%%%%%

%%%%%%%%%%%%%%%%%%%%%%%%%%%%%%%%%%%%%%%%%%%%%%%%%%%%%%%%%%%%%%%%%%%%
\subsection{Parameters}
\label{Sec:Parameter}
%%%%%%%%%%%%%%%%%%%%%%%%%%%%%%%%%%%%%%%%%%%%%%%%%%%%%%%%%%%%%%%%%%%%

%%%%%%
%%%%%% parameter:qqbar+qq
%%%%%%
\begin{table*}
	\centering
	\caption{Parameters of the $q\bar{q}$ pairs for mesons and of the $qq$ pairs for baryons~\cite{Weng:2018mmf} (in units of $\text{MeV}$).}
	\label{table:parameter:qqbar+qq}
	\begin{tabular}{lcccccccccccc}
		\toprule[1pt]
		\toprule[1pt]
		Parameter&$m_{n\bar{n}}^{m}$&$m_{n\bar{s}}^{m}$&$m_{s\bar{s}}^{m}$&$m_{n\bar{c}}^{m}$&$m_{s\bar{c}}^{m}$&$m_{c\bar{c}}^{m}$&$m_{n\bar{b}}^{m}$&$m_{s\bar{b}}^{m}$&$m_{c\bar{b}}^{m}$&$m_{b\bar{b}}^{m}$\\
		Value&$615.95$&$794.22$&$936.40$&$1973.22$&$2076.14$&$3068.53$&$5313.35$&$5403.25$&$6322.27$&$9444.97$\\
		%\midrule[1pt]
		Parameter&$v_{n\bar{n}}^{m}$&$v_{n\bar{s}}^{m}$&$v_{s\bar{s}}^{m}$&$v_{n\bar{c}}^{m}$&$v_{s\bar{c}}^{m}$&$v_{c\bar{c}}^{m}$&$v_{n\bar{b}}^{m}$&$v_{s\bar{b}}^{m}$&$v_{c\bar{b}}^{m}$&$v_{b\bar{b}}^{m}$\\
		Value&$477.92$&$298.57$&$249.18$&$106.01$&$107.87$&$85.12$&$33.89$&$36.43$&$47.18$&$45.98$\\
		\midrule[1pt]
		Parameter&$m_{nn}^{b}$&$m_{ns}^{b}$&$m_{ss}^{b}$&$m_{nc}^{b}$&$m_{sc}^{b}$&$m_{cc}^{b}$&$m_{nb}^{b}$&$m_{sb}^{b}$&$m_{cb}^{b}$&$m_{b{b}}^{b}$\\
		Value&$724.85$&$906.65$&$1049.36$&$2079.96$&$2183.68$&$3171.51$&$5412.25$&$5494.80$&$6416.07$&$9529.57$\\
		%\midrule[1pt]
		Parameter&$v_{n{n}}^{b}$&$v_{n{s}}^{b}$&$v_{ss}^{b}$&$v_{n{c}}^{b}$&$v_{s{c}}^{b}$&$v_{c{c}}^{b}$&$v_{n{b}}^{b}$&$v_{s{b}}^{b}$&$v_{c{b}}^{b}$&$v_{b{b}}^{b}$\\
		Value&$305.34$&$212.75$&$195.30$&$62.81$&$70.63$&$56.75$&$19.92$&$8.47$&$31.45$&$30.65$\\
		\bottomrule[1pt]
		\bottomrule[1pt]
	\end{tabular}
\end{table*}
%
%%%%%%
%%%%%% parameter:delta-m_q
%%%%%%
\begin{table}
\centering
\caption{Values of the difference ${\delta}\tilde{m}_{q}^{bm}=m_{q}^{b}-m_{q}^{m}+\frac{4}{3}\delta{a}_{q}^{bm}$~\cite{Weng:2018mmf} (in units of $\text{MeV}$).}
\label{table:parameter:delta-m_q}
\begin{tabular}{lcccccccccccc}
\toprule[1pt]
\toprule[1pt]
&$\delta\tilde{m}_{n}^{bm}$&~&$\delta\tilde{m}_{s}^{bm}$&~&$\delta\tilde{m}_{c}^{bm}$&~&$\delta\tilde{m}_{b}^{bm}$\\
%%%
\midrule[1pt]
%%%
Value&$54.94$&&$56.48$&&$51.49$&&$42.30$\\
%%%
\bottomrule[1pt]
\bottomrule[1pt]
\end{tabular}
\end{table}
%
%%%%%%
%%%%%% parameter:tetraquark
%%%%%%
\begin{table}
\centering
\caption{Possible choices of tetraquark parameters.}
\label{table:parameter:tetraquark}
\begin{tabular}{lcccccccccccc}
\toprule[1pt]
\toprule[1pt]
&~&$m_{q_{i}q_{j}}^{t}$&~&$m_{q_{i}\bar{q}_{j}}^{t}$&~&$v_{q_{i}q_{j}}^{t}$&~&$v_{q_{i}\bar{q}_{j}}^{t}$\\
%%%
\midrule[1pt]
%%%%
Scheme~I&&$m_{q_{i}q_{j}}^{b}$&&$m_{q_{i}\bar{q}_{j}}^{m}$&&$v_{q_{i}q_{j}}^{b}$&&$v_{q_{i}\bar{q}_{j}}^{m}$\\
%%%
Scheme~II&&$m_{q_{i}q_{j}}^{b}$&&$m_{q_{i}{q}_{j}}^{b}$&&$v_{q_{i}q_{j}}^{b}$&&$v_{q_{i}\bar{q}_{j}}^{m}$\\
%%%
Scheme~III&&$m_{q_{i}q_{j}}^{b}$&&$m_{q_{i}\bar{q}_{j}}^{m}$&&$v_{q_{i}q_{j}}^{b}$&&$v_{q_{i}{q}_{j}}^{b}$\\
%%%
Scheme~IV&&$m_{q_{i}q_{j}}^{b}$&&$m_{q_{i}{q}_{j}}^{b}$&&$v_{q_{i}q_{j}}^{b}$&&$v_{q_{i}{q}_{j}}^{b}$\\
%%%
\bottomrule[1pt]
\bottomrule[1pt]
\end{tabular}
\end{table}

To calculate the tetraquark masses, one needs to estimate the parameters $\{m_{ij}^{t},v_{ij}^{t}\}$.
In Ref.~\cite{Weng:2018mmf} we used the meson and baryon masses to extract the parameters $\{m_{q_{1}\bar{q}_{2}}^{m},v_{q_{1}\bar{q}_{2}}^{m}\}$ and $\{m_{q_{1}q_{2}}^{b},v_{q_{1}q_{2}}^{b}\}$.
The baryon parameters $\{m_{Q_{1}Q_{2}}^{b},v_{Q_{1}Q_{2}}^{b}\}$ between two heavy quarks cannot be fitted from baryons because of the lack of experimental data.
For this reason, we adopted the assumptions
\begin{equation}
{\delta}a_{q_{1}q_{2}}^{bm}
{\equiv}
a_{q_{1}q_{2}}^{b}-a_{q_{1}\bar{q}_{2}}^{m}
{\approx}
0
\end{equation}
and
\begin{equation}
R_{q_{1}q_{2}}^{bm}
{\equiv}
v_{q_{1}q_{2}}^{b}/v_{q_{1}\bar{q}_{2}}^{m}
=
2/3\pm0.30
\end{equation}
to estimate them from the meson parameters $\{m_{Q_{1}\bar{Q}_{2}}^{m},v_{Q_{1}\bar{Q}_{2}}^{m}\}$.
The resulting parameters are listed in Table~\ref{table:parameter:qqbar+qq}.
Since the CM interaction strength $v_{ij}$'s are inversely proportional to the quark masses, the meson parameters $\{v_{c\bar{c}},v_{c\bar{b}},v_{b\bar{b}}\}$ between heavy flavors are quite small.
Thus the large uncertainty of the ratio $R_{q_{1}q_{2}}^{bm}$ does no have much effects on the baryon parameters $\{v_{cc},v_{cb},v_{bb}\}$ and the mass spectrum of the doubly heavy tetraquarks.
As shown in Ref.~\cite{Weng:2018mmf}, the introduction of the first assumption makes the difference $\delta{m}_{q_{1}q_{2}}^{bm}{\equiv}m_{q_{1}q_{2}}^{b}-m_{q_{1}\bar{q}_{2}}^{m}$ separable over the two quarks
\begin{equation}
\delta{m}_{q_{1}q_{2}}^{bm}
%%%
{\approx}
{\delta}{m}_{q_{1}}^{bm}
+
{\delta}{m}_{q_{2}}^{bm}
\end{equation}
where ${\delta}{m}_{q}^{bm}{\equiv}m_{q}^{b}-m_{q}^{m}$ is the difference of the effective quark mass extracted from the baryon and meson.
In this way, the ten $\delta{m}_{q_{1}q_{2}}^{bm}$'s reduce to four ${\delta}{m}_{q}^{bm}$'s.
Actually, such property can be achieved by a weaker assumption.
Namely, we assume that the difference $a_{q_{1}q_{2}}^{b}-a_{q_{1}\bar{q}_{2}}^{m}$ is separable over the two quarks
\begin{equation}
a_{q_{1}q_{2}}^{b}
-
a_{q_{1}\bar{q}_{2}}^{m}
%%%
{\approx}
\delta{a}_{q_{1}}^{bm}
+
\delta{a}_{q_{2}}^{bm}
\end{equation}
Then we have
\begin{equation}
\delta{m}_{q_{1}q_{2}}^{bm}
%%%
{\approx}
{\delta}\tilde{m}_{q_{1}}^{bm}
+
{\delta}\tilde{m}_{q_{2}}^{bm}
\end{equation}
where
\begin{equation}
{\delta}\tilde{m}_{q}^{bm}
{\equiv}
m_{q}^{b}-m_{q}^{m}+\frac{4}{3}\delta{a}_{q}^{bm}
=
\delta{m}_{q}^{bm}+\frac{4}{3}\delta{a}_{q}^{bm}
\end{equation}
which includes the quark mass difference and the differences between color interactions.
We again reduce the ten $\delta{m}_{q_{1}q_{2}}^{bm}$'s into four degrees of freedom.
All results are unchanged except that we reinterpret the $\delta{m}_{q}^{bm}$ of Ref.~\cite{Weng:2018mmf} as ${\delta}\tilde{m}_{q}^{bm}$ (see Table~\ref{table:parameter:delta-m_q} or Table~VI of Ref.~\cite{Weng:2018mmf}).

Now we consider the tetraquarks.
In Ref.~\cite{Weng:2020jao}, We used the following scheme to estimate the masses of the fully heavy tetraquarks
\begin{equation}
m_{q_{i}q_{j}}^{t}
\approx
m_{q_{i}q_{j}}^{b}\,,
\end{equation}
\begin{equation}
	m_{q_{i}\bar{q}_{j}}^{t}
	\approx
	m_{q_{i}\bar{q}_{j}}^{m}\,,
\end{equation}
\begin{equation}
	v_{q_{i}q_{j}}^{t}
	\approx
	v_{q_{i}q_{j}}^{b}\,,
\end{equation}
\begin{equation}
v_{q_{i}\bar{q}_{j}}^{t}
\approx
v_{q_{i}\bar{q}_{j}}^{m}\,.
\end{equation}
Within this scheme, we found that the ground states of the fully heavy tetraquarks are dominated by color-sextet configurations, which is consistent with the dynamical calculations~\cite{Wang:2019rdo,Deng:2020iqw}.
Nonetheless, this scheme ignores the difference of the spatial configurations between the tetraquarks and the normal hadrons, which will evidently cause large uncertainties~\cite{Cui:2006mp,Zhao:2014qva,Deng:2020iqw}.
To appreciate the uncertainty, we introduce three additional schemes for comparison (see Table~\ref{table:parameter:tetraquark}).
The scheme~III (IV) differs from the scheme~I (II) by
\begin{equation}
v_{q_{i}\bar{q}_{j}}^{t}
\approx
v_{q_{i}\bar{q}_{j}}^{m}
\quad
\Longrightarrow
\quad
v_{q_{i}\bar{q}_{j}}^{t}
\approx
v_{q_{i}{q}_{j}}^{b}\,.
\end{equation}
Due to the smallness of $v_{qQ}^{b}$ and $v_{q\bar{Q}}^{m}$, the results in scheme~I (II) are very similar to those in scheme~III (IV).
Thus we will focus on the scheme~I and scheme~II.
%

%%%%%%%%%%%%%%%%%%%%%%%%%%%%%%%%%%%%%%%%%%%%%%%%%%%%%%%%%%%%%%%%%%%%
\subsection{The $nn\bar{Q}\bar{Q}$ systems}
\label{sec:nnQQ}
%%%%%%%%%%%%%%%%%%%%%%%%%%%%%%%%%%%%%%%%%%%%%%%%%%%%%%%%%%%%%%%%%%%%

%%%
%%% mass:nncc+nnbb+nncb:I+II
%%%
\begin{table*}%[htbp]
	\centering
	\caption{Masses and eigenvectors of the $nn\bar{c}\bar{c}$, $nn\bar{b}\bar{b}$ and $nn\bar{c}\bar{b}$ tetraquarks. The masses are all in units of MeV.}
	\label{table:mass:nncc+nnbb+nncb:I+II}
	\begin{tabular}{ccccccc}
		\toprule[1pt]
		\toprule[1pt]
		\multirow{2}{*}{System}&\multirow{2}{*}{$J^{P}$}&\multicolumn{2}{c}{Scheme~I}&\multicolumn{2}{c}{Scheme~II}\\
		\cmidrule(lr){3-4}
		\cmidrule(lr){5-6}
		&&Mass&Eigenvector&Mass&Eigenvector\\
		\midrule[1pt]
		%%%
		$(nn\bar{c}\bar{c})^{I=1}$&$0^{+}$
		&$3833.2$&$\{0.515,0.857\}$
		&$3969.2$&$\{0.350,0.937\}$\\
		&
		&$4127.4$&$\{0.857,-0.515\}$
		&$4364.9$&$\{0.937,-0.350\}$\\
		%%%
		&$1^{+}$
		&$3946.4$&$\{1\}$
		&$4053.2$&$\{1\}$\\
		%%%
		&$2^{+}$
		&$4017.1$&$\{1\}$
		&$4123.8$&$\{1\}$\\
		%%%
		%%%
		$(nn\bar{c}\bar{c})^{I=0}$&$1^{+}$
		&$3749.8$&$\{0.354,-0.935\}$
		&$3868.7$&$\{0.212,-0.977\}$\\
		&
		&$3976.1$&$\{0.935,0.354\}$
		&$4230.8$&$\{0.977,0.212\}$\\
		%%%
		\midrule[1pt]
		%%%
		$(nn\bar{b}\bar{b})^{I=1}$&$0^{+}$
		&$10468.8$&$\{0.123,0.992\}$
		&$10569.3$&$\{0.086,0.996\}$\\
		&
		&$10808.9$&$\{0.992,-0.123\}$
		&$11054.6$&$\{0.996,-0.086\}$\\
		%%%
		&$1^{+}$
		&$10485.3$&$\{1\}$
		&$10584.2$&$\{1\}$\\
		%%%
		&$2^{+}$
		&$10507.9$&$\{1\}$
		&$10606.8$&$\{1\}$\\
		%%%
		%%%
		$(nn\bar{b}\bar{b})^{I=0}$&$1^{+}$
		&$10291.6$&$\{0.058,-0.998\}$
		&$10390.9$&$\{0.043,-0.999\}$\\
		&
		&$10703.4$&$\{0.998,0.058\}$
		&$10950.3$&$\{0.999,0.043\}$\\
		%%%
		\midrule[1pt]
		%%%
		$(nn\bar{c}\bar{b})^{I=1}$&$0^{+}$
		&$7189.5$&$\{0.366,0.931\}$
		&$7305.6$&$\{0.232,0.973\}$\\
		&
		&$7440.9$&$\{0.931,-0.366\}$
		&$7684.7$&$\{0.973,-0.232\}$\\
		%%%
		&$1^{+}$
		&$7211.0$&$\{-0.311,-0.648,0.696\}$
		&$7322.5$&$\{-0.180,-0.687,0.704\}$\\
		&
		&$7264.2$&$\{-0.048,0.742,0.669\}$
		&$7367.3$&$\{-0.029,0.719,0.694\}$\\
		&
		&$7417.0$&$\{0.949,-0.175,0.262\}$
		&$7665.1$&$\{0.983,-0.104,0.150\}$\\
		%%%
		&$2^{+}$
		&$7293.2$&$\{1\}$
		&$7396.0$&$\{1\}$\\
		%%%
		%%%
		$(nn\bar{c}\bar{b})^{I=0}$&$0^{+}$
		&$7003.4$&$\{0.440,0.898\}$
		&$7124.6$&$\{0.266,0.964\}$\\
		&
		&$7220.3$&$\{0.898,-0.440\}$
		&$7459.0$&$\{0.964,-0.266\}$\\
		%%%
		&$1^{+}$
		&$7046.2$&$\{0.228,-0.219,0.949\}$
		&$7158.0$&$\{0.122,-0.133,0.984\}$\\
		&
		&$7232.9$&$\{0.899,-0.327,-0.292\}$
		&$7482.4$&$\{0.910,-0.381,-0.165\}$\\
		&
		&$7329.3$&$\{-0.374,-0.919,-0.122\}$
		&$7584.9$&$\{-0.397,-0.915,-0.074\}$\\
		%%%
		&$2^{+}$
		&$7353.2$&$\{1\}$
		&$7610.3$&$\{1\}$\\
		%%%
		\bottomrule[1pt]
		\bottomrule[1pt]
	\end{tabular}
\end{table*}
%
%%%
%%% eigenvector:nncc:I+II
%%%
\begin{table*}%[htbp]
	\centering
	\caption{The eigenvectors of the $nn\bar{c}\bar{c}$ tetraquark states in the $n\bar{c}{\otimes}n\bar{c}$ configuration. The masses are all in units of MeV.}
	\label{table:eigenvector:nncc:I+II}
	\begin{tabular}{cccccccccccccc}
		\toprule[1pt]
		\toprule[1pt]
		\multirow{2}{*}{System}&\multirow{2}{*}{$J^{P}$}&\multicolumn{5}{c}{Scheme~I}&\multicolumn{5}{c}{Scheme~II}\\
		\cmidrule(lr){3-7}
		\cmidrule(lr){8-12}
		&
		&Mass&$\bar{D}^{*}\bar{D}^{*}$&$\bar{D}^{*}\bar{D}$&$\bar{D}\bar{D}^{*}$&$\bar{D}\bar{D}$
		&Mass&$\bar{D}^{*}\bar{D}^{*}$&$\bar{D}^{*}\bar{D}$&$\bar{D}\bar{D}^{*}$&$\bar{D}\bar{D}$\\
		\midrule[1pt]
		%%%
		$(nn\bar{c}\bar{c})^{I=1}$&$0^{+}$
		&$3833.2$&$0.116$&&&$0.639$
		&$3969.2$&$-0.023$&&&$0.611$\\
		&
		&$4127.4$&$0.755$&&&$0.093$
		&$4364.9$&$0.763$&&&$0.207$\\
		%%%
		&$1^{+}$
		&$3946.4$&$0$&$0.408$&$0.408$&
		&$4053.2$&$0$&$0.408$&$0.408$\\
		%%%
		&$2^{+}$
		&$4017.1$&$0.577$&&&
		&$4123.8$&$0.577$\\
		%%%
		%%%
		$(nn\bar{c}\bar{c})^{I=0}$&$1^{+}$
		&$3749.8$&$-0.177$&$0.415$&$-0.415$&
		&$3868.7$&$-0.277$&$0.369$&$-0.369$\\
		&
		&$3976.1$&$0.685$&$0.280$&$-0.280$&
		&$4230.8$&$0.651$&$0.338$&$-0.338$\\
		%%%
		\bottomrule[1pt]
		\bottomrule[1pt]
	\end{tabular}
\end{table*}
%
%%%
%%% kc_i^2:nncc:I+II
%%%
\begin{table*}%[htbp]
	\centering
	\caption{The values of $k\cdot|c_{i}|^2$ for the $nn\bar{c}\bar{c}$ tetraquarks (in unit of MeV).}
	\label{table:kc_i^2:nncc:I+II}
	\begin{tabular}{cccccccccc}
		\toprule[1pt]
		\toprule[1pt]
		\multirow{2}{*}{System}&\multirow{2}{*}{$J^{P}$}&\multicolumn{4}{c}{Scheme~I}&\multicolumn{4}{c}{Scheme~II}\\
		\cmidrule(lr){3-6}
		\cmidrule(lr){7-10}
		&
		&Mass&$\bar{D}^{*}\bar{D}^{*}$&$\bar{D}\bar{D}^{*}$&$\bar{D}\bar{D}$
		&Mass&$\bar{D}^{*}\bar{D}^{*}$&$\bar{D}\bar{D}^{*}$&$\bar{D}\bar{D}$\\
		\midrule[1pt]
		%%%
		$(nn\bar{c}\bar{c})^{I=1}$&$0^{+}$
		&$3833.2$
		&$\times$&&$176.4$
		&$3969.2$
		&$\times$&&$251.3$\\
		&%%
		&$4127.4$
		&$270.0$&&$7.6$
		&$4364.9$
		&$497.6$&&$48.5$\\
		%%%
		&$1^{+}$
		&$3946.4$&&$61.9$&
		&$4053.2$
		&&$98.8$\\
		%%%
		&$2^{+}$
		&$4017.1$&$\times$&&
		&$4123.8$
		&$155.3$\\
		%%%
		%%%
		$(nn\bar{c}\bar{c})^{I=0}$&$1^{+}$
		&$3749.8$&$\times$&$\times$&
		&$3868.7$
		&$\times$&$\times$\\
		&
		&$3976.1$&$\times$&$34.7$&
		&$4230.8$
		&$281.1$&$96.8$\\
		%%%
		\bottomrule[1pt]
		\bottomrule[1pt]
	\end{tabular}
\end{table*}
%
%%%
%%% R:nncc:I+II
%%%
\begin{table*}%[htbp]
	\centering
	\caption{The partial width ratios for the $nn\bar{c}\bar{c}$ tetraquarks. For each state, we choose one mode as the reference channel, and the partial width ratios of the other channels are calculated relative to this channel. The masses are all in unit of MeV.}
	\label{table:R:nncc:I+II}
	\begin{tabular}{cccccccccc}
		\toprule[1pt]
		\toprule[1pt]
		\multirow{2}{*}{System}&\multirow{2}{*}{$J^{P}$}&\multicolumn{4}{c}{Scheme~I}&\multicolumn{4}{c}{Scheme~II}\\
		\cmidrule(lr){3-6}
		\cmidrule(lr){7-10}
		&
		&Mass&$\bar{D}^{*}\bar{D}^{*}$&$\bar{D}\bar{D}^{*}$&$\bar{D}\bar{D}$
		&Mass&$\bar{D}^{*}\bar{D}^{*}$&$\bar{D}\bar{D}^{*}$&$\bar{D}\bar{D}$\\
		\midrule[1pt]
		%%%
		$(nn\bar{c}\bar{c})^{I=1}$&$0^{+}$
		&$3833.2$&$\times$&&$1$
		&$3969.2$
		&$\times$&&$1$\\
		&%%
		&$4127.4$&$35.7$&&$1$
		&$4364.9$
		&$10.3$&&$1$\\
		%%%
		&$1^{+}$
		&$3946.4$&&$1$&
		&$4053.2$
		&&$1$\\
		%%%
		&$2^{+}$
		&$4017.1$&$\times$&&
		&$4123.8$
		&$1$\\
		%%%
		%%%
		$(nn\bar{c}\bar{c})^{I=0}$&$1^{+}$
		&$3749.8$&$\times$&$\times$&
		&$3868.7$
		&$\times$&$\times$\\
		&
		&$3976.1$&$\times$&$1$&
		&$4230.8$
		&$1.5$&$1$\\
		%%%
		\bottomrule[1pt]
		\bottomrule[1pt]
	\end{tabular}
\end{table*}
%
%%%
%%% eigenvector:nnbb:I+II
%%%
\begin{table*}%[htbp]
	\centering
	\caption{The eigenvectors of the $nn\bar{b}\bar{b}$ tetraquark states in the $n\bar{b}{\otimes}n\bar{b}$ configuration. The masses are all in units of MeV.}
	\label{table:eigenvector:nnbb:I+II}
	\begin{tabular}{cccccccccccccccc}
		\toprule[1pt]
		\toprule[1pt]
		\multirow{2}{*}{System}&\multirow{2}{*}{$J^{P}$}&\multicolumn{5}{c}{Scheme~I}&\multicolumn{5}{c}{Scheme~II}\\
		\cmidrule(lr){3-7}
		\cmidrule(lr){8-12}
		&
		&Mass&$B^{*}B^{*}$&$B^{*}B$&$BB^{*}$&$BB$
		&Mass&$B^{*}B^{*}$&$B^{*}B$&$BB^{*}$&$BB$\\
		\midrule[1pt]
		%%%
		$(nn\bar{b}\bar{b})^{I=1}$&$0^{+}$
		&$10468.8$&$-0.200$&&&$0.546$
		&$10569.3$&$-0.227$&&&$0.533$\\
		&
		&$10808.9$&$0.737$&&&$0.344$
		&$11054.6$&$0.729$&&&$0.364$\\
		%%%
		&$1^{+}$
		&$10485.3$&$0$&$0.408$&$0.408$&
		&$10584.2$&$0$&$0.408$&$0.408$\\
		%%%
		&$2^{+}$
		&$10507.9$&$0.577$&&&
		&$10606.8$&$0.577$\\
		%%%
		%%%
		$(nn\bar{b}\bar{b})^{I=0}$&$1^{+}$
		&$10291.6$&$-0.374$&$0.312$&$-0.312$&
		&$10390.9$&$-0.383$&$0.306$&$-0.306$\\
		&
		&$10703.4$&$0.600$&$0.391$&$-0.391$&
		&$10950.3$&$0.594$&$0.395$&$-0.395$\\
		%%%
		\bottomrule[1pt]
		\bottomrule[1pt]
	\end{tabular}
\end{table*}
%
%%%
%%% kc_i^2:nnbb:I+II
%%%
\begin{table*}%[htbp]
	\centering
	\caption{The values of $k\cdot|c_{i}|^2$ for the $nn\bar{b}\bar{b}$ tetraquarks (in unit of MeV).}
	\label{table:kc_i^2:nnbb:I+II}
	\begin{tabular}{cccccccccc}
		\toprule[1pt]
		\toprule[1pt]
		\multirow{2}{*}{System}&\multirow{2}{*}{$J^{P}$}&\multicolumn{4}{c}{Scheme~I}&\multicolumn{4}{c}{Scheme~II}\\
		\cmidrule(lr){3-6}
		\cmidrule(lr){7-10}
		&
		&Mass&$B^{*}B^{*}$&$BB^{*}$&$BB$
		&Mass&$B^{*}B^{*}$&$BB^{*}$&$BB$\\
		\midrule[1pt]
		%%%
		$(nn\bar{b}\bar{b})^{I=1}$&$0^{+}$
		&$10468.8$
		&$\times$&&$\times$
		&$10569.3$
		&$\times$&&$66.5$\\
		&
		&$10808.9$
		&$503.0$&&$136.5$
		&$11054.6$
		&$788.7$&&$216.6$\\
		%%%
		&$1^{+}$
		&$10485.3$
		&&$\times$&
		&$10584.2$
		&&$\times$\\
		%%%
		&$2^{+}$
		&$10507.9$
		&$\times$&&
		&$10606.8$
		&$\times$\\
		%%%
		%%%
		$(nn\bar{b}\bar{b})^{I=0}$&$1^{+}$
		&$10291.6$
		&$\times$&$\times$&
		&$10390.9$
		&$\times$&$\times$\\
		&
		&$10703.4$
		&$193.6$&$111.0$&
		&$10950.3$
		&$450.3$&$213.6$\\
		%%%
		\bottomrule[1pt]
		\bottomrule[1pt]
	\end{tabular}
\end{table*}
%
%%%
%%% R:nnbb:I+II
%%%
\begin{table*}%[htbp]
	\centering
	\caption{The partial width ratios for the $nn\bar{b}\bar{b}$ tetraquarks. For each state, we choose one mode as the reference channel, and the partial width ratios of the other channels are calculated relative to this channel. The masses are all in unit of MeV.}
	\label{table:R:nnbb:I+II}
	\begin{tabular}{cccccccccc}
		\toprule[1pt]
		\toprule[1pt]
		\multirow{2}{*}{System}&\multirow{2}{*}{$J^{P}$}&\multicolumn{4}{c}{Scheme~I}&\multicolumn{4}{c}{Scheme~II}\\
		\cmidrule(lr){3-6}
		\cmidrule(lr){7-10}
		&
		&Mass&$B^{*}B^{*}$&$BB^{*}$&$BB$
		&Mass&$B^{*}B^{*}$&$BB^{*}$&$BB$\\
		\midrule[1pt]
		%%%
		$(nn\bar{b}\bar{b})^{I=1}$&$0^{+}$
		&$10468.8$
		&$\times$&&$\times$
		&$10569.3$
		&$\times$&&$1$\\
		&
		&$10808.9$&$3.7$&&$1$
		&$11054.6$
		&$3.6$&&$1$\\
		%%%
		&$1^{+}$
		&$10485.3$
		&&$\times$&
		&$10584.2$
		&&$\times$\\
		%%%
		&$2^{+}$
		&$10507.9$
		&$\times$&&
		&$10606.8$
		&$\times$\\
		%%%
		%%%
		$(nn\bar{b}\bar{b})^{I=0}$&$1^{+}$
		&$10291.6$
		&$\times$&$\times$&
		&$10390.9$
		&$\times$&$\times$\\
		&
		&$10703.4$
		&$0.9$&$1$&
		&$10950.3$
		&$1.1$&$1$\\
		%%%
		\bottomrule[1pt]
		\bottomrule[1pt]
	\end{tabular}
\end{table*}
%
%%%
%%% eigenvector:nncb:I+II
%%%
\begin{table*}%[htbp]
	\centering
	\caption{The eigenvectors of the $nn\bar{c}\bar{b}$ tetraquark states in the $n\bar{c}{\otimes}n\bar{b}$ configuration. The masses are all in units of MeV.}
	\label{table:eigenvector:nncb:I+II}
	\begin{tabular}{ccccccccccccccc}
		\toprule[1pt]
		\toprule[1pt]
		\multirow{2}{*}{System}&\multirow{2}{*}{$J^{P}$}&\multicolumn{5}{c}{Scheme~I}&\multicolumn{5}{c}{Scheme~II}\\
		\cmidrule(lr){3-7}
		\cmidrule(lr){8-12}
		&
		&Mass&$\bar{D}^{*}B^{*}$&$\bar{D}^{*}B$&$\bar{D}B^{*}$&$\bar{D}B$
		&Mass&$\bar{D}^{*}B^{*}$&$\bar{D}^{*}B$&$\bar{D}B^{*}$&$\bar{D}B$\\
		\midrule[1pt]
		%%%
		$(nn\bar{c}\bar{b})^{I=1}$&$0^{+}$
		&$7189.5$&$-0.010$&&&$0.615$
		&$7305.6$&$-0.116$&&&$0.581$\\
		&
		&$7440.9$&$0.764$&&&$0.197$
		&$7684.7$&$0.755$&&&$0.281$\\
		%%%
		&$1^{+}$
		&$7211.0$&$0.104$&$0.063$&$-0.592$&
		&$7322.5$&$0.184$&$-0.004$&$-0.557$\\
		&
		&$7264.2$&$0.245$&$0.515$&$0.090$&
		&$7367.3$&$0.266$&$0.506$&$0.081$\\
		&
		&$7417.0$&$0.655$&$-0.383$&$0.241$&
		&$7665.1$&$0.629$&$-0.401$&$0.316$\\
		%%%
		&$2^{+}$
		&$7293.2$&$0.577$&&&
		&$7396.0$&$0.577$\\
		%%%
		%%%
		$(nn\bar{c}\bar{b})^{I=0}$&$0^{+}$
		&$7003.4$&$0.269$&&&$0.570$
		&$7124.6$&$0.374$&&&$0.466$\\
		&
		&$7220.3$&$-0.587$&&&$0.508$
		&$7459.0$&$-0.526$&&&$0.605$\\
		%%%
		&$1^{+}$
		&$7046.2$&$0.261$&$-0.231$&$0.495$&
		&$7158.0$&$0.325$&$-0.268$&$0.409$\\
		&
		&$7232.9$&$-0.308$&$0.469$&$0.568$&
		&$7482.4$&$-0.287$&$0.417$&$0.633$\\
		&
		&$7329.3$&$-0.580$&$-0.556$&$0.124$&
		&$7584.9$&$-0.559$&$-0.581$&$0.123$\\
		%%%
		&$2^{+}$
		&$7353.2$&$0.817$&&&
		&$7610.3$&$0.817$\\
		%%%
		\bottomrule[1pt]
		\bottomrule[1pt]
	\end{tabular}
\end{table*}
%
%%%
%%% kc_i^2:nncb:I+II
%%%
\begin{table*}%[htbp]
	\centering
	\caption{The values of $k\cdot|c_{i}|^2$ for the $nn\bar{c}\bar{b}$ tetraquarks (in unit of MeV).}
	\label{table:kc_i^2:nncb:I+II}
	\begin{tabular}{ccccccccccccc}
		\toprule[1pt]
		\toprule[1pt]
		\multirow{2}{*}{System}&\multirow{2}{*}{$J^{P}$}&\multicolumn{5}{c}{Scheme~I}&\multicolumn{5}{c}{Scheme~II}\\
		\cmidrule(lr){3-7}
		\cmidrule(lr){8-12}
		&
		&Mass&$\bar{D}^{*}B^{*}$&$\bar{D}^{*}B$&$\bar{D}B^{*}$&$\bar{D}B$
		&Mass&$\bar{D}^{*}B^{*}$&$\bar{D}^{*}B$&$\bar{D}B^{*}$&$\bar{D}B$\\
		\midrule[1pt]
		%%%
		$(nn\bar{c}\bar{b})^{I=1}$&$0^{+}$
		&$7189.5$
		&$\times$&&&$130.4$
		&$7305.6$
		&$\times$&&&$226.3$\\
		&
		&$7440.9$
		&$329.3$&&&$35.6$
		&$7684.7$
		&$590.5$&&&$99.9$\\
		%%%
		&$1^{+}$
		&$7211.0$
		&$\times$&$\times$&$80.7$&
		&$7322.5$
		&$\times$&$0.004$&$188.4$\\
		&
		&$7264.2$
		&$\times$&$\times$&$3.7$&
		&$7367.3$
		&$22.4$&$123.6$&$4.7$\\
		&
		&$7417.0$
		&$213.2$&$90.8$&$46.4$&
		&$7665.1$
		&$397.6$&$172.5$&$117.9$\\
		%%%
		&$2^{+}$
		&$7293.2$
		&$\times$&&&
		&$7396.0$
		&$143.3$\\
		%%%
		%%%
		$(nn\bar{c}\bar{b})^{I=0}$&$0^{+}$
		&$7003.4$
		&$\times$&&&$\times$
		&$7124.6$
		&$\times$&&&$\times$\\
		&
		&$7220.3$
		&$\times$&&&$117.0$
		&$7459.0$
		&$169.3$&&&$347.6$\\
		%%%
		&$1^{+}$
		&$7046.2$
		&$\times$&$\times$&$\times$&
		&$7158.0$
		&$\times$&$\times$&$\times$\\
		&
		&$7232.9$
		&$\times$&$\times$&$109.1$&
		&$7482.4$
		&$55.0$&$132.7$&$367.1$\\
		&
		&$7329.3$
		&$\times$&$107.5$&$9.6$&
		&$7584.9$
		&$271.9$&$320.1$&$16.2$\\
		%%%
		&$2^{+}$
		&$7353.2$
		&$161.3$&&&
		&$7610.3$
		&$610.5$\\
		%%%
		\bottomrule[1pt]
		\bottomrule[1pt]
	\end{tabular}
\end{table*}
%
%%%
%%% R:nncb:I+II
%%%
\begin{table*}%[htbp]
	\centering
	\caption{The partial width ratios for the $nn\bar{c}\bar{b}$ tetraquarks. For each state, we choose one mode as the reference channel, and the partial width ratios of the other channels are calculated relative to this channel. The masses are all in unit of MeV.}
	\label{table:R:nncb:I+II}
	\begin{tabular}{ccccccccccccccc}
		\toprule[1pt]
		\toprule[1pt]
		\multirow{2}{*}{System}&\multirow{2}{*}{$J^{P}$}&\multicolumn{5}{c}{Scheme~I}&\multicolumn{5}{c}{Scheme~II}\\
		\cmidrule(lr){3-7}
		\cmidrule(lr){8-12}
		&
		&Mass&$\bar{D}^{*}B^{*}$&$\bar{D}^{*}B$&$\bar{D}B^{*}$&$\bar{D}B$
		&Mass&$\bar{D}^{*}B^{*}$&$\bar{D}^{*}B$&$\bar{D}B^{*}$&$\bar{D}B$\\
		\midrule[1pt]
		%%%
		$(nn\bar{c}\bar{b})^{I=1}$&$0^{+}$
		&$7189.5$
		&$\times$&&&$1$
		&$7305.6$
		&$\times$&&&$1$\\
		&
		&$7440.9$
		&$9.2$&&&$1$
		&$7684.7$
		&$5.9$&&&$1$\\
		%%%
		&$1^{+}$
		&$7211.0$
		&$\times$&$\times$&$1$&
		&$7322.5$
		&$\times$&$0.00002$&$1$\\
		&
		&$7264.2$
		&$\times$&$\times$&$1$&
		&$7367.3$
		&$4.8$&$26.5$&$1$\\
		&
		&$7417.0$
		&$4.6$&$2.0$&$1$&
		&$7665.1$
		&$3.4$&$1.5$&$1$\\
		%%%
		&$2^{+}$
		&$7293.2$
		&$\times$&&&
		&$7396.0$
		&$1$\\
		%%%
		%%%
		$(nn\bar{c}\bar{b})^{I=0}$&$0^{+}$
		&$7003.4$
		&$\times$&&&$\times$
		&$7124.6$
		&$\times$&&&$\times$\\
		&
		&$7220.3$
		&$\times$&&&$1$
		&$7459.0$
		&$0.5$&&&$1$\\
		%%%
		&$1^{+}$
		&$7046.2$
		&$\times$&$\times$&$\times$&
		&$7158.0$
		&$\times$&$\times$&$\times$\\
		&
		&$7232.9$
		&$\times$&$\times$&$1$&
		&$7482.4$
		&$0.1$&$0.4$&$1$\\
		&
		&$7329.3$
		&$\times$&$11.2$&$1$&
		&$7584.9$
		&$16.8$&$19.7$&$1$\\
		%%%
		&$2^{+}$
		&$7353.2$
		&$1$&&&
		&$7610.3$
		&$1$\\
		%%%
		\bottomrule[1pt]
		\bottomrule[1pt]
	\end{tabular}
\end{table*}
%
%%%
%%% mass:nncc+nnbb:I+II
%%%
\begin{figure*}%[!h]
	\begin{tabular}{ccc}
		\includegraphics[width=450pt]{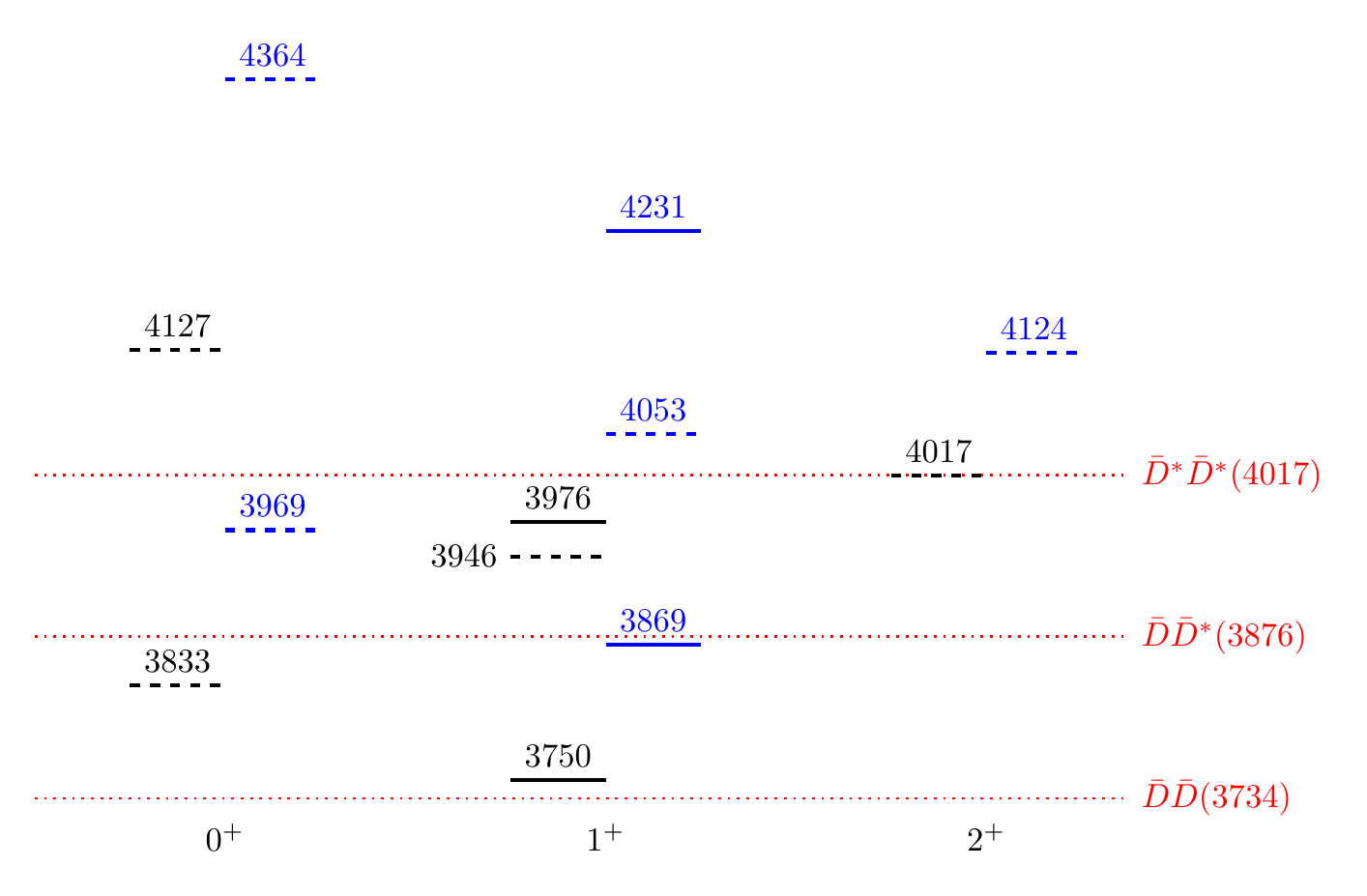}\\
		(a) $nn\bar{c}\bar{c}$ states\\
		&&\\
		\includegraphics[width=450pt]{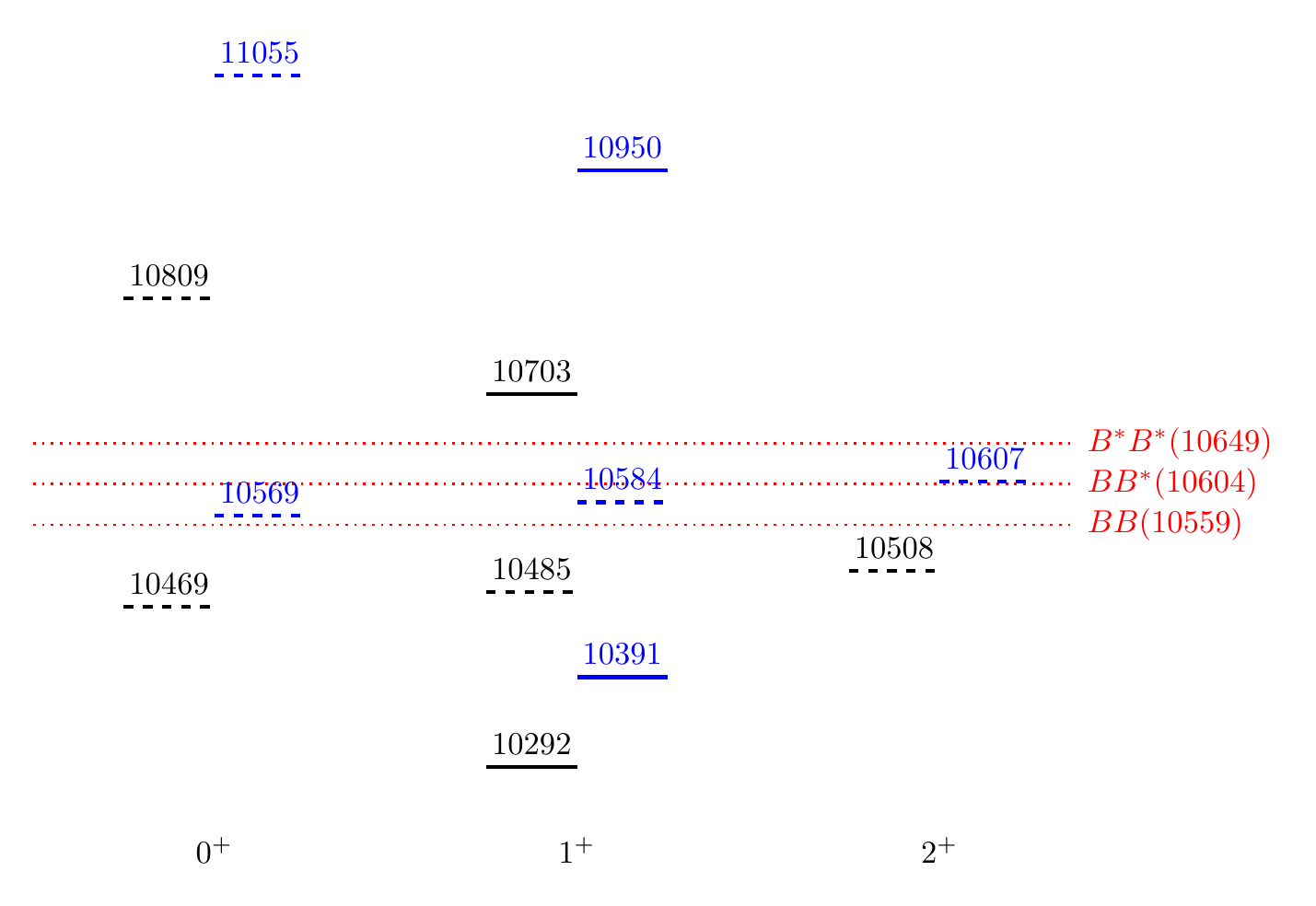}\\
		(b) $nn\bar{b}\bar{b}$ states\\
	\end{tabular}
	\caption{Mass spectra of the $I=0$ (solid) and $I=1$ (dashed) $nn\bar{c}\bar{c}$ and $nn\bar{b}\bar{b}$ tetraquark states in scheme~I (black) and scheme~II (blue). The dotted lines indicate various meson-meson thresholds. The masses are all in units of MeV.}
	\label{fig:nncc+nnbb}
\end{figure*}
%
%%%
%%% mass:nncb+:I+II
%%%
\begin{figure*}%[!h]
	\begin{tabular}{ccc}
		\includegraphics[width=450pt]{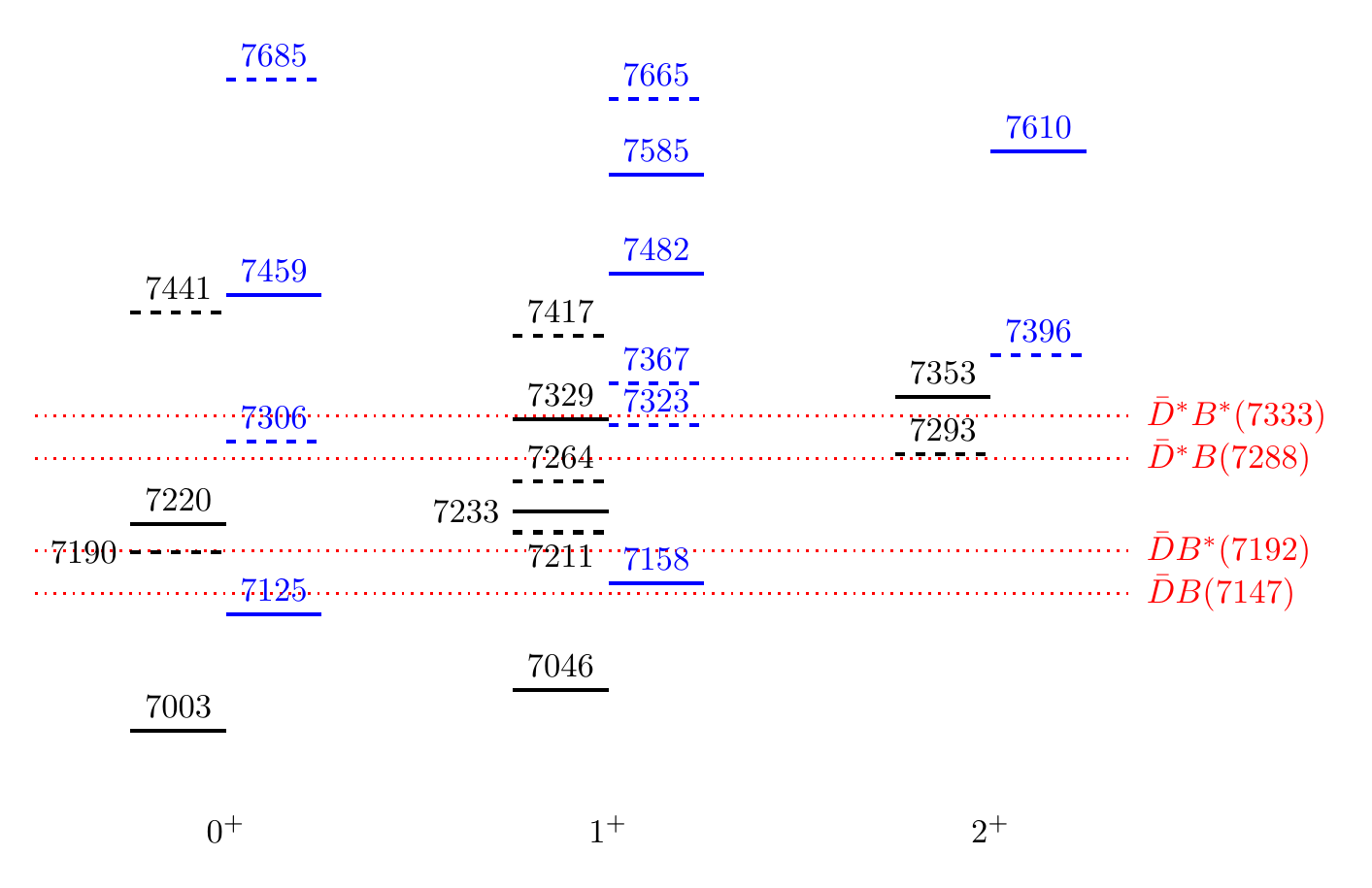}\\
	\end{tabular}
	\caption{Mass spectra of the $I=0$ (solid) and $I=1$ (dashed) $nn\bar{c}\bar{b}$ tetraquark states in scheme~I (black) and scheme~II (blue). The dotted lines indicate various meson-meson thresholds. The masses are all in units of MeV.}
	\label{fig:nncb}
\end{figure*}
%

%%%%%%%%%%%%%%%%%%%%%%%%%%%%%%%%%%%%%%%%%%%%%%%%%%%%%%%%%%%%%%%%%%%%
\subsubsection{The $nn\bar{c}\bar{c}$ and $nn\bar{b}\bar{b}$ tetraquarks}
\label{sec:nncc+nnbb}
%%%%%%%%%%%%%%%%%%%%%%%%%%%%%%%%%%%%%%%%%%%%%%%%%%%%%%%%%%%%%%%%%%%%

Inserting the parameters into the Hamiltonian, we can determine the tetraquark masses.
The masses and eigenvectors of the $nn\bar{Q}\bar{Q}$ tetraquarks are listed in Table~\ref{table:mass:nncc+nnbb+nncb:I+II}.
Here, we assume that the $\textrm{SU}(2)$ flavor symmetry is exact and denote $u$, $d$ quarks collectively as $n$.
In the following, we will use $T_{i}(nn\bar{Q}\bar{Q},m,I,J^{P})$ to represent the $nn\bar{Q}\bar{Q}$ tetraquarks, where the subscript $i$ denotes the particular scheme of the parameters.
%
%The $nn\bar{Q}\bar{Q}$ tetraquarks can be of isospin $I=1$ and of isospin $I=0$.
%
In Figs~\ref{fig:nncc+nnbb}--\ref{fig:nncb}, we plot the relative position of the $nn\bar{Q}\bar{Q}$ tetraquarks and their meson-meson thresholds.
%

%We first consider the results of scheme~I.
%
We first consider the $nn\bar{c}\bar{c}$ tetraquarks.
%
%We first consider the $nn\bar{c}\bar{c}$ and $nn\bar{b}\bar{b}$ tetraquarks.
%
The quantum number of its lightest state is $IJ^{P}=01^{+}$, namely the $T_{I}(nn\bar{c}\bar{c},3749.8,0,1^{+})$ or $T_{II}(nn\bar{c}\bar{c},3868.7,0,1^{+})$ state.
The other isoscalar state is $T_{I}(nn\bar{c}\bar{c},3976.1,0,1^{+})$ or $T_{II}(nn\bar{c}\bar{c},4230.8,0,1^{+})$.
We find that the scheme~II always gives larger masses than the scheme~I.
The reason is that the two schemes choose different value of $m_{q_{i}\bar{q}_{j}}^{t}$, which results in different values of the color interaction.
More precisely, the difference of the color interaction between the two schemes is
\begin{align}
\Delta{H}_{\text{C}}
%%%
={}&H_{\text{C}}^{II}-H_{\text{C}}^{I}
\notag\\
%%%
={}&
-\frac{3}{4}
\sum_{i<j}
%\sum_{1{\leq}i<j{\leq}4}
\left(m_{ij}^{II}-m_{ij}^{I}\right)
V_{ij}^{\text{C}}
\notag\\
%%%
={}&
-\frac{3}{4}
\sum_{i\leq2}
\sum_{j>2}
\delta{m}_{ij}^{bm}
V_{ij}^{\text{C}}
\notag\\
%%%
\approx{}&
-\frac{3}{4}
\sum_{i\leq2}
\sum_{j>2}
\left(\delta\tilde{m}_{i}^{bm}+\delta\tilde{m}_{j}^{bm}\right)
\bm{F}_{i}\cdot\bm{F}_{j}
\notag\\
%%%
={}&
\frac{3}{8}
\left(
\sum_{i}
\delta\tilde{m}_{i}^{bm}
\right)
\cdot
\left(
\sum_{i\leq2}
\bm{F}_{i}
\right)^{2}
\end{align}
where in the last line we have ignored the terms proportional to $\sum_{i}\bm{F}_{i}$.
Note that both $\Ket{(q_{1}q_{2})^{6_{c}}(\bar{q}_{3}\bar{q}_{4})^{\bar{6}_{c}}}$ and $\Ket{(q_{1}q_{2})^{\bar{3}_{c}}(\bar{q}_{3}\bar{q}_{4})^{3_{c}}}$ are eigenstates of $(\bm{F}_{1}+\bm{F}_{2})^{2}$, with eigenvalues $10/3$ and $4/3$ respectively.
%
%In other words, $\Delta{H}_{\text{C}}$ pushes up the color-sextet and color-triplet configurations by $\frac{5}{4}\sum_{i}\delta\tilde{m}_{i}^{bm}$ and $\frac{1}{2}\sum_{i}\delta\tilde{m}_{i}^{bm}$ respectively.
%
In other words,
\begin{equation}\label{eqn:HC:diff}
\Braket{\Delta{H}_{\text{C}}}
=
\left\{
\begin{split}
&\frac{5}{4}\sum_{i}\delta\tilde{m}_{i}^{bm}\,,\text{~for~}\ket{(q_{1}q_{2})^{6_{c}}(\bar{q}_{3}\bar{q}_{4})^{\bar{6}_{c}}}\,,\\
&\frac{1}{2}\sum_{i}\delta\tilde{m}_{i}^{bm}\,,\text{~for~}\ket{(q_{1}q_{2})^{\bar{3}_{c}}(\bar{q}_{3}\bar{q}_{4})^{3_{c}}}\,.
\end{split}
\right.
\end{equation}
For the $nn\bar{c}\bar{c}$ system, $\sum_{i}\delta\tilde{m}_{i}^{bm}=212.9~\text{MeV}$.
The ground state $T_{I}(nn\bar{c}\bar{c},3749.8,0,1^{+})$ is dominated by the color-triplet configuration, and its mass is increased by about $118.9~\text{MeV}$.
While the mass of the color-sextet configuration dominated state $T_{I}(nn\bar{c}\bar{c},3976.1,0,1^{+})$ is increased by $254.7~\text{MeV}$.
The deviation from Eq.~\eqref{eqn:HC:diff} is caused by the color mixing.
In the isovector sector, we have four tetraquark states.
They are all above the corresponding $S$-wave decay channels.
It is interesting to note that the $T_{II}(nn\bar{c}\bar{c},3686.7,0,1^{+})$ in scheme~II is quite close to the newly observed $T_{cc}^{+}$ state.
%

%~\\

%\textcolor{red}{$T_{cc}$}

%~\\

The $nn\bar{b}\bar{b}$ tetraquarks is very similar to the $nn\bar{c}\bar{c}$ tetraquarks.
Its lightest state also have quantum number $IJ^{P}=01^{+}$, namely the $T_{I}(nn\bar{b}\bar{b},10291.6,0,1^{+})$ or $T_{II}(nn\bar{b}\bar{b},10390.9,0,1^{+})$.
In both schemes, this state lies below the $BB$ threshold and is stable against strong decays.
In scheme~I, the $T_{I}(nn\bar{b}\bar{b},10468.9,1,0^{+})$, $T_{I}(nn\bar{b}\bar{b},10485.3,1,1^{+})$ and $T_{I}(nn\bar{b}\bar{b},10507.9,1,2^{+})$ also lie below the the $BB$ threshold.
But they are not stable in scheme~II.
Thus we cannot draw a definite conclusion.

Besides the masses, the eigenvectors also help understand the nature of the tetraquarks.
Within the four possible quantum numbers, the $IJ^{P}=10^{+}$ one and the $IJ^{P}=01^{+}$ one are of particular interest because they both have two possible color configurations, namely the color-sextet $\ket{(qq)^{6_{c}}{\otimes}(\bar{Q}\bar{Q})^{\bar{6}_{c}}}$ and color-triplet $\ket{(qq)^{\bar{3}_{c}}{\otimes}(\bar{Q}\bar{Q})^{{3}_{c}}}$.
For simplicity, we denote them as $6_{c}{\otimes}\bar{6}_{c}$ and $\bar{3}_{c}{\otimes}{3}_{c}$.
As pointed out by Wang~\etal~\cite{Wang:2019rdo}, there are two competing effects in determining whether the $6_{c}{\otimes}\bar{6}_{c}$ or $\bar{3}_{c}{\otimes}{3}_{c}$ dominates the tetraquark's ground state.
In the one-gluon-exchange (OGE) model, the color interactions in color-triplet diquark are attractive, while those in color-sextet diquark are repulsive.
On the other hand, the attractions between $6_{c}$ diquark and $\bar{6}_{c}$ anti-diquark 
and between the $\bar{3}_{c}{\otimes}{3}_{c}$ counterpart are both attractive, and the former one is much stronger.
The authors of Refs.~\cite{Wang:2019rdo,Deng:2020iqw,Weng:2020jao} found that the color-sextet configuration has more net attractions for most fully heavy tetraquarks.
Thus the ground states contain more color-sextet components than the color-triplet one.
The only exception is the $cc\bar{b}\bar{b}$ tetraquark in model~II of Ref~\cite{Wang:2019rdo}, whose ground state has $53\%$ of the $\bar{3}_{c}{\otimes}3_{c}$ component.
It is also interesting to note that, when the mass ratio between quarks and antiquarks deviates from one, the color-triplet configuration becomes more important in the ground states.
For example, Ref.~\cite{Weng:2020jao} found that the $T(bb\bar{b}\bar{b},18836.1,0^{++})$ and $T(cc\bar{c}\bar{c},6044.9,0^{++})$ have $18.5\%$ and $30.5\%$ of the $\bar{3}_{c}{\otimes}3_{c}$ components, while the $T(cc\bar{b}\bar{b},12596.3,0^{++})$ has $48.4\%$.
This tendency also exists in the doubly heavy tetraquarks.
As shown in Table~\ref{table:mass:nncc+nnbb+nncb:I+II}, the $\bar{3}_{c}{\otimes}3_{c}$ components become dominant in ground states of the $nn\bar{Q}\bar{Q}$ tetraquarks.
This phenomenon can also be explained by the color interaction Hamiltonian,
\begin{align}
&\Braket{H_{\text{C}}\left(nn\bar{Q}\bar{Q}\right)}
\notag\\
%%%
={}&
%%%
-\frac{3}{4}
\Braket{
m_{nn}^{t}V_{12}^{\text{C}}
+
m_{QQ}^{t}V_{34}^{\text{C}}
+
m_{n\bar{Q}}^{t}
\left(V_{13}^{\text{C}}+V_{24}^{\text{C}}+V_{14}^{\text{C}}+V_{23}^{\text{C}}\right)
}
\notag\\
%%%
={}&
%%%
-\frac{3}{4}
\Braket{
m_{n\bar{Q}}^{t}
\sum_{i<j}V^{\text{C}}_{ij}
+
2\delta{m}
\left(V_{12}^{\text{C}}+V_{34}^{\text{C}}\right)
}
\notag\\
%%%
={}&
2m_{n\bar{Q}}^{t}
-
\frac{3}{2}
\delta{m}
\Braket{V_{12}^{\text{C}}+V_{34}^{\text{C}}}
\notag\\
%%%
={}&
2m_{n\bar{Q}}^{t}
+
\delta{m}
\begin{pmatrix}
%-1_{6_{c}\otimes\bar{6}_{c}}&0\\
%0&+2_{\bar{3}_{c}\otimes3_{c}}
-1&0\\
0&+2
\end{pmatrix}
\end{align}
where we have expanded the Hamiltonian in the bases  \{$\ket{\left(n_1n_2\right)_{0}^{6}\left(\bar{Q}_{3}\bar{Q}_{4}\right)_{0}^{\bar{6}}}_{0}$,$\ket{\left(n_1n_2\right)_{1}^{\bar{3}}\left(\bar{Q}_{3}\bar{Q}_{4}\right)_{1}^{3}}_{0}$\} in the last line and
\begin{equation}
\delta{m}
=
\frac{1}{2}
\left(\frac{m_{nn}^{t}+m_{QQ}^{t}}{2}-m_{n\bar{Q}}^{t}\right)
\end{equation}
Taking scheme~I as an example, we have
\begin{align}
\delta{m}\left(nn\bar{c}\bar{c}\right)=-12.52~\text{MeV}\,,\\
\delta{m}\left(nn\bar{b}\bar{b}\right)=-93.07~\text{MeV}\,,
\end{align}
while for the fully heavy tetraquarks
\begin{align}
\delta{m}\left(bb\bar{b}\bar{b}\right)=+42.30~\text{MeV}\,,\\
\delta{m}\left(cc\bar{c}\bar{c}\right)=+51.49~\text{MeV}\,,\\
\delta{m}\left(cc\bar{b}\bar{b}\right)=+15.15~\text{MeV}\,.
\end{align}
As the ratios $m_{\bar{q}}/m_{q}$'s increase, the $\bar{3}_{c}{\otimes}3_{c}$ components become more important in the ground states.

Another interesting conclusion from the Hamiltonian is that the color interaction does not mix the $6_{c}{\otimes}\bar{6}_{c}$ and $\bar{3}_{c}{\otimes}3_{c}$ configurations.
Actually, this conclusion applies for all $S$-wave tetraquarks with $q_{1}=q_{2}$ or $\bar{q}_{3}=\bar{q}_{4}$.
Let's consider the matrix element of color interaction $\Braket{\alpha|H_{\text{CE}}|\beta}$.
Note that the color interaction is independent of the spin operator, and thus is a rank-0 tensor in the $q_{1}q_{2}$ spin space.
Its matrix elements over different $q_{1}q_{2}$ spin states always vanish.
If $q_{1}=q_{2}$, the Pauli principle further renders the matrix elements vanish unless the bases $\alpha$ and $\beta$ possess the same color symmetry over $q_{1}q_{2}$.
The same argument works for $\bar{q}_{3}\bar{q}_{4}$ as well.
In summary, the color interaction does no mix the $6_{c}{\otimes}\bar{6}_{c}$ and $\bar{3}_{c}{\otimes}3_{c}$ color configurations if $q_{1}=q_{2}$ or $\bar{q}_{3}=\bar{q}_{4}$.
%
%Taken the $J^{P}=0^{+}$ states as example.
%
%The matrix element of color interaction $\braket{H_{\text{CE}}}$ should be a $2\times2$ block diagonal matrix composed of two $2\times2$ matrices, while the first one is over the bases $\{\alpha_{1}^{0},\alpha_{4}^{0}\}$ and the second one is over the bases $\{\alpha_{2}^{0},\alpha_{3}^{0}\}$.
%

%~\\

%\textcolor{red}{$H_{\text{CM}}$}

%~\\

Next we consider their decay properties.
For the decays of the tetraquark states in this work, the $(k/m)^2$'s are all of $\mathcal{O}(10^{-2})$ or even smaller.
All higher wave decays are suppressed.
% except for the $S$-wave decays.
%
Thus we will only consider the $S$-wave decays in this work.
First we transform the wave function of $nn\bar{Q}\bar{Q}$ tetraquarks into the $n\bar{Q}{\otimes}n\bar{Q}$ configuration.
Then we can calculate the $k\cdot|c_{i}|^2$'s and partial decay width ratios.
The corresponding results are listed in Tables~\ref{table:eigenvector:nncc:I+II}--\ref{table:R:nnbb:I+II}.
%
%When calculating the final state momentum $k$, we have used the theoretical meson masses for consistency.
%
Note that the two schemes give very similar results, we will mainly focus on the scheme~I in the following.
In the isovector sector, the $nn\bar{c}\bar{c}$ tetraquarks are mostly above the $S$-wave decay channels, thus are wide states.
Depending on the schemes, the $J^{P}=2^{+}$ state may lie on or above the $\bar{D}^{*}\bar{D}^{*}$ threshold.
Namely the $T_{I}(nn\bar{c}\bar{c},4017.1,1,2^{+})$ or $T_{II}(nn\bar{c}\bar{c},4123.8,1,2^{+})$.
A firm conclusion requires more detailed studies.
The $T_{I}(nn\bar{c}\bar{c},4127.4,1,0^{+})$ can decay into both $\bar{D}\bar{D}$ and $\bar{D}^{*}\bar{D}^{*}$ channels, with partial decay width ratio
\begin{equation}
\Gamma_{\bar{D}^{*}\bar{D}^{*}}:\Gamma_{\bar{D}\bar{D}}
\sim37.5\,.
\end{equation}
Thus the $\bar{D}^{*}\bar{D}^{*}$ mode is dominant.
%
%In the isoscalar sector, $T_{I}(nn\bar{c}\bar{c},3749.8,0,1^{+})$ is narrow since it lies below the $\bar{D}\bar{D}^{*}$ threshold.
%
In the isoscalar sector, the $T_{I}(nn\bar{c}\bar{c},3749.8,0,1^{+})$ lies below the $\bar{D}\bar{D}^{*}$ threshold, thus it is a narrow state.
However, this state lies above the $\bar{D}\bar{D}$ threshold, so it can decay radiatively into the $\bar{D}\bar{D}\gamma$ final states.
The $T_{I}(nn\bar{c}\bar{c},3976.1,0,1^{+})$ is wide because it lies above the $\bar{D}\bar{D}^{*}$ and $\bar{D}^{*}\bar{D}^{*}$ thresholds.
The $T_{I}(nn\bar{b}\bar{b},10808.9,1,0^{+})$ and $T_{I}(nn\bar{b}\bar{b},10703.4,0,1^{+})$ lie above all the corresponding $S$-wave decay channel.
Their partial decay width ratios are
\begin{equation}
\frac{\Gamma[T_{I}(nn\bar{b}\bar{b},10808.9,1,0^{+}){\rightarrow}B^{*}B^{*}]}{\Gamma[T_{I}(nn\bar{b}\bar{b},10808.9,1,0^{+}){\rightarrow}BB]}
\sim3.7
\end{equation}
and
\begin{equation}
\frac{\Gamma[T_{I}(nn\bar{b}\bar{b},10703.4,0,1^{+}){\rightarrow}B^{*}B^{*}]}{\Gamma[T_{I}(nn\bar{b}\bar{b},10703.4,0,1^{+}){\rightarrow}BB]}
\sim0.9
\end{equation}
respectively.
Other $nn\bar{b}\bar{b}$ tetraquarks are all narrow states.
%

%%%%%%%%%%%%%%%%%%%%%%%%%%%%%%%%%%%%%%%%%%%%%%%%%%%%%%%%%%%%%%%%%%%%
\subsubsection{The $nn\bar{c}\bar{b}$ tetraquark}
\label{sec:nncb}
%%%%%%%%%%%%%%%%%%%%%%%%%%%%%%%%%%%%%%%%%%%%%%%%%%%%%%%%%%%%%%%%%%%%

Next we consider the $nn\bar{c}\bar{b}$ tetraquark.
We list the masses and eigenvectors of these states in Table~\ref{table:mass:nncc+nnbb+nncb:I+II}.
Their relative position and possible decay channels are plotted in Fig.~\ref{fig:nncb}.
There are two possible stable $nn\bar{c}\bar{b}$ tetraquark states.
The first state is $T_{I}(nn\bar{c}\bar{b},7003.4,0,0^{+})$, which lies below the $\bar{D}B$ threshold by more than 100~MeV.
Even in scheme~II, this state is about 20~MeV below the threshold.
The second state is $T_{I}(nn\bar{c}\bar{b},7046.2,0,1^{+})$.
It is about 100~MeV lighter than the $\bar{D}B$ threshold.
However, this state lies above the threshold in scheme~II.
Nonetheless, it lies below its $S$-wave decay mode $\bar{D}B^{*}$, thus should be a narrow state.

Since the two antiquarks do not have to obey the Pauli principle, we have much bigger number of states than the $nn\bar{c}\bar{c}$/$nn\bar{b}\bar{b}$ cases.
For each isospin, we have two $0^{+}$ states, three $1^{+}$ states and one $2^{+}$ state.
From Table~\ref{table:mass:nncc+nnbb+nncb:I+II}, we see that for each possible quantum number, the lower mass states are dominated by color-triplet configurations, while the color-sextet configurations are more important in the higher mass states.
For example, the two stable states $T_{I}(nn\bar{c}\bar{b},7003.4,0,0^{+})$ and $T_{I}(nn\bar{c}\bar{b},7046.2,0,1^{+})$ have $80.6\%$ and $90.0\%$ of $\bar{3}_{c}{\otimes}3_{c}$ components respectively.
This can be explained by the color interaction
\begin{equation}
\Braket{H_{\text{C}}\left(nn\bar{c}\bar{b}\right)}%_{J}
%%%
=
m_{n\bar{c}}
+
m_{n\bar{b}}
-
\frac{3}{2}
\delta{m}'
\Braket{V^{\text{C}}_{12}+V^{\text{C}}_{34}}%_{J}
\end{equation}
where
\begin{align}
\delta{m}'
={}&
\frac{1}{4}
\left(m_{nn}+m_{cb}-m_{n\bar{c}}-m_{n\bar{b}}\right)
\notag\\
%%%
={}&
-36.41~\text{MeV}
\end{align}
Note that both $6_{c}\otimes\bar{6}_{c}$ and $\bar{3}_{c}\otimes3_{c}$ configurations are eigenstates of $V^{\text{C}}_{12}+V^{\text{C}}_{34}$, with eigenvalues $2/3$ and $-4/3$ respectively.
The negative value of $\delta{m}'$ indicates that the color interaction favors the $\bar{3}_{c}\otimes3_{c}$ configuration.

In Table~\ref{table:eigenvector:nncb:I+II}, we transform the $nn\bar{c}\bar{b}$ tetraquarks into the $n\bar{c}{\otimes}n\bar{b}$ configuration.
Then we calculate the values of $k\cdot|c_{i}|^2$ and relative partial decay widths, as shown in Tables~\ref{table:kc_i^2:nncb:I+II}--\ref{table:R:nncb:I+II}.
Besides the two stable states discussed above, two heavier isoscalar states $T_{I}(nn\bar{c}\bar{b},7220.3,0,0^{+})$ and $T_{I}(nn\bar{c}\bar{b},7232.9,0,1^{+})$ are above the $\bar{D}B^{*}$, while $T_{I}(nn\bar{c}\bar{b},7329.3,0,1^{+})$ can decay into both $\bar{D}^{*}B$ and $\bar{D}B^{*}$ modes, with relative width
\begin{equation}
\Gamma_{\bar{D}^{*}B}:\Gamma_{\bar{D}B^{*}}
\sim11.2:1
\end{equation}
In the isovector sector, the lower $0^{+}$ state can only decay into $\bar{D}B$ mode, while the higher one can also decay into $\bar{D}^{*}B^{*}$ mode.
All $1^{+}$ states can decay into $\bar{D}B^{*}$ in $S$-wave, while only the highest one can decay into $\bar{D}^{*}B$ and $\bar{D}^{*}B^{*}$ modes, with partial decay rates
\begin{equation}
\Gamma_{\bar{D}^{*}B^{*}}:\Gamma_{\bar{D}^{*}B}:\Gamma_{\bar{D}B}
\sim4.6:2.0:1
\end{equation}
There is no doubt that the current results rely on the mass estimation.
In scheme~II, the higher masses allow the states to have more decay modes.
Yet we find that the partial decay width ratios are quite stable in the two schemes.
%

%%%%%%%%%%%%%%%%%%%%%%%%%%%%%%%%%%%%%%%%%%%%%%%%%%%%%%%%%%%%%%%%%%%%
\subsection{The $ss\bar{Q}\bar{Q}$ systems}
\label{sec:ssQQ}
%%%%%%%%%%%%%%%%%%%%%%%%%%%%%%%%%%%%%%%%%%%%%%%%%%%%%%%%%%%%%%%%%%%%

%%%
%%% mass:sscc+ssbb+sscb:I+II
%%%
\begin{table*}%[htbp]
	\centering
	\caption{Masses and eigenvectors of the $ss\bar{c}\bar{c}$, $ss\bar{b}\bar{b}$ and $ss\bar{c}\bar{b}$ tetraquarks. The masses are all in units of MeV.}
	\label{table:mass:sscc+ssbb+sscb:I+II}
	\begin{tabular}{ccccccc}
		\toprule[1pt]
		\toprule[1pt]
		\multirow{2}{*}{System}&\multirow{2}{*}{$J^{P}$}&\multicolumn{2}{c}{Scheme~I}&\multicolumn{2}{c}{Scheme~II}\\
		\cmidrule(lr){3-4}
		\cmidrule(lr){5-6}
		&&Mass&Eigenvector&Mass&Eigenvector\\
		\midrule[1pt]
		%%%
		$ss\bar{c}\bar{c}$&$0^{+}$
		&$4043.7$&$\{0.650,0.760\}$
		&$4199.1$&$\{0.442,0.897\}$\\
		&
		&$4311.1$&$\{0.760,-0.650\}$
		&$4532.1$&$\{0.897,-0.442\}$\\
		%%%
		&$1^{+}$
		&$4192.6$&$\{1\}$
		&$4300.2$&$\{1\}$\\
		%%%
		&$2^{+}$
		&$4264.5$&$\{1\}$
		&$4372.1$&$\{1\}$\\
		%%%
		\midrule[1pt]
		%%%
		$ss\bar{b}\bar{b}$&$0^{+}$
		&$10697.1$&$\{0.196,0.981\}$
		&$10792.1$&$\{0.124,0.992\}$\\
		&
		&$10928.8$&$\{0.981,-0.196\}$
		&$11154.3$&$\{0.992,-0.124\}$\\
		%%%
		&$1^{+}$
		&$10718.2$&$\{1\}$
		&$10809.8$&$\{1\}$\\
		%%%
		&$2^{+}$
		&$10742.5$&$\{1\}$
		&$10834.1$&$\{1\}$\\
		%%%
		\midrule[1pt]
		%%%
		$ss\bar{c}\bar{b}$&$0^{+}$
		&$7404.4$&$\{0.547,0.837\}$
		&$7531.3$&$\{0.325,0.946\}$\\
		&
		&$7597.3$&$\{0.837,-0.547\}$
		&$7818.8$&$\{0.946,-0.325\}$\\
		%%%
		&$1^{+}$
		&$7431.8$&$\{-0.537,-0.551,0.639\}$
		&$7553.6$&$\{-0.265,-0.662,0.701\}$\\
		&
		&$7503.3$&$\{-0.096,0.792,0.603\}$
		&$7603.5$&$\{-0.047,0.735,0.677\}$\\
		&
		&$7569.4$&$\{0.838,-0.263,0.478\}$
		&$7795.3$&$\{0.963,-0.146,0.226\}$\\
		%%%
		&$2^{+}$
		&$7534.3$&$\{1\}$
		&$7633.8$&$\{1\}$\\
		%%%
		\bottomrule[1pt]
		\bottomrule[1pt]
	\end{tabular}
\end{table*}
%
%%%
%%% eigenvector:sscc:I+II
%%%
\begin{table*}%[htbp]
	\centering
	\caption{The eigenvectors of the $ss\bar{c}\bar{c}$ tetraquark states in the $s\bar{c}{\otimes}s\bar{c}$ configuration. The masses are all in units of MeV.}
	\label{table:eigenvector:sscc:I+II}
	\begin{tabular}{cccccccccccccc}
		\toprule[1pt]
		\toprule[1pt]
		\multirow{2}{*}{System}&\multirow{2}{*}{$J^{P}$}&\multicolumn{5}{c}{Scheme~I}&\multicolumn{5}{c}{Scheme~II}\\
		\cmidrule(lr){3-7}
		\cmidrule(lr){8-12}
		&
		&Mass&$\bar{D}_{s}^{*}\bar{D}_{s}^{*}$&$\bar{D}_{s}^{*}\bar{D}_{s}$&$\bar{D}_{s}\bar{D}_{s}^{*}$&$\bar{D}_{s}\bar{D}_{s}$
		&Mass&$\bar{D}_{s}^{*}\bar{D}_{s}^{*}$&$\bar{D}_{s}^{*}\bar{D}_{s}$&$\bar{D}_{s}\bar{D}_{s}^{*}$&$\bar{D}_{s}\bar{D}_{s}$\\
		\midrule[1pt]
		%%%
		$ss\bar{c}\bar{c}$&$0^{+}$
		&$4043.7$&$0.240$&&&$0.645$
		&$4199.1$&$0.054$&&&$0.629$\\
		&
		&$4311.1$&$0.725$&&&$-0.015$
		&$4532.1$&$0.762$&&&$0.145$\\
		%%%
		&$1^{+}$
		&$4192.6$&$0$&$0.408$&$0.408$&
		&$4300.2$&$0$&$0.408$&$0.408$\\
		%%%
		&$2^{+}$
		&$4264.5$&$0.577$&&&
		&$4372.1$&$0.577$\\
		%%%
		\bottomrule[1pt]
		\bottomrule[1pt]
	\end{tabular}
\end{table*}
%
%%%
%%% kc_i^2:sscc:I+II
%%%
\begin{table*}%[htbp]
	\centering
	\caption{The values of $k\cdot|c_{i}|^2$ for the $ss\bar{c}\bar{c}$ tetraquarks (in unit of MeV).}
	\label{table:kc_i^2:sscc:I+II}
	\begin{tabular}{cccccccccc}
		\toprule[1pt]
		\toprule[1pt]
		\multirow{2}{*}{System}&\multirow{2}{*}{$J^{P}$}&\multicolumn{4}{c}{Scheme~I}&\multicolumn{4}{c}{Scheme~II}\\
		\cmidrule(lr){3-6}
		\cmidrule(lr){7-10}
		&
		&Mass&$\bar{D}_{s}^{*}\bar{D}_{s}^{*}$&$\bar{D}_{s}\bar{D}_{s}^{*}$&$\bar{D}_{s}\bar{D}_{s}$
		&Mass&$\bar{D}_{s}^{*}\bar{D}_{s}^{*}$&$\bar{D}_{s}\bar{D}_{s}^{*}$&$\bar{D}_{s}\bar{D}_{s}$\\
		\midrule[1pt]
		%%%
		$ss\bar{c}\bar{c}$&$0^{+}$
		&$4043.7$
		&$\times$&&$192.5$
		&$4199.1$
		&$\times$&&$289.1$\\
		&
		&$4311.1$
		&$226.4$&&$0.2$
		&$4532.1$
		&$476.6$&&$23.6$\\
		%%%
		&$1^{+}$
		&$4192.6$
		&&$80.3$&
		&$4300.2$
		&&$113.0$\\
		%%%
		&$2^{+}$
		&$4264.5$
		&$97.5$&&
		&$4372.1$
		&$187.9$\\
		%%%
		\bottomrule[1pt]
		\bottomrule[1pt]
	\end{tabular}
\end{table*}
%
%%%
%%% R:sscc:I+II
%%%
\begin{table*}%[htbp]
	\centering
	\caption{The partial width ratios for the $ss\bar{c}\bar{c}$ tetraquarks. For each state, we choose one mode as the reference channel, and the partial width ratios of the other channels are calculated relative to this channel. The masses are all in unit of MeV.}
	\label{table:R:sscc:I+II}
	\begin{tabular}{cccccccccc}
		\toprule[1pt]
		\toprule[1pt]
		\multirow{2}{*}{System}&\multirow{2}{*}{$J^{P}$}&\multicolumn{4}{c}{Scheme~I}&\multicolumn{4}{c}{Scheme~II}\\
		\cmidrule(lr){3-6}
		\cmidrule(lr){7-10}
		&
		&Mass&$\bar{D}_{s}^{*}\bar{D}_{s}^{*}$&$\bar{D}_{s}\bar{D}_{s}^{*}$&$\bar{D}_{s}\bar{D}_{s}$
		&Mass&$\bar{D}_{s}^{*}\bar{D}_{s}^{*}$&$\bar{D}_{s}\bar{D}_{s}^{*}$&$\bar{D}_{s}\bar{D}_{s}$\\
		\midrule[1pt]
		%%%
		$ss\bar{c}\bar{c}$&$0^{+}$
		&$4043.7$
		&$\times$&&$1$
		&$4199.1$
		&$\times$&&$1$\\
		&
		&$4311.1$
		&$1185.7$&&$1$
		&$4532.1$
		&$20.2$&&$1$\\
		%%%
		&$1^{+}$
		&$4192.6$
		&&$1$&
		&$4300.2$
		&&$1$\\
		%%%
		&$2^{+}$
		&$4264.5$
		&$1$&&
		&$4372.1$
		&$1$\\
		%%%
		\bottomrule[1pt]
		\bottomrule[1pt]
	\end{tabular}
\end{table*}
%
%%%
%%% eigenvector:ssbb:I+II
%%%
\begin{table*}%[htbp]
	\centering
	\caption{The eigenvectors of the $ss\bar{b}\bar{b}$ tetraquark states in the $s\bar{b}{\otimes}s\bar{b}$ configuration. The masses are all in units of MeV.}
	\label{table:eigenvector:ssbb:I+II}
	\begin{tabular}{cccccccccccccc}
		\toprule[1pt]
		\toprule[1pt]
		\multirow{2}{*}{System}&\multirow{2}{*}{$J^{P}$}&\multicolumn{5}{c}{Scheme~I}&\multicolumn{5}{c}{Scheme~II}\\
		\cmidrule(lr){3-7}
		\cmidrule(lr){8-12}
		&
		&Mass&$B_{s}^{*}B_{s}^{*}$&$B_{s}^{*}B_{s}$&$B_{s}B_{s}^{*}$&$B_{s}B_{s}$
		&Mass&$B_{s}^{*}B_{s}^{*}$&$B_{s}^{*}B_{s}$&$B_{s}B_{s}^{*}$&$B_{s}B_{s}$\\
		\midrule[1pt]
		%%%
		$ss\bar{b}\bar{b}$&$0^{+}$
		&$10697.1$&$-0.144$&&&$0.570$
		&$10792.1$&$-0.199$&&&$0.547$\\
		&
		&$10928.8$&$0.750$&&&$0.302$
		&$11154.3$&$0.737$&&&$0.343$\\
		%%%
		&$1^{+}$
		&$10718.2$&$0$&$0.408$&$0.408$&
		&$10809.8$&$0$&$0.408$&$0.408$\\
		%%%
		&$2^{+}$
		&$10742.5$&$0.577$&&&
		&$10834.1$&$0.577$\\
		%%%
		\bottomrule[1pt]
		\bottomrule[1pt]
	\end{tabular}
\end{table*}
%
%%%
%%% kc_i^2:ssbb:I+II
%%%
\begin{table*}%[htbp]
	\centering
	\caption{The values of $k\cdot|c_{i}|^2$ for the $ss\bar{b}\bar{b}$ tetraquarks (in unit of MeV).}
	\label{table:kc_i^2:ssbb:I+II}
	\begin{tabular}{cccccccccc}
		\toprule[1pt]
		\toprule[1pt]
		\multirow{2}{*}{System}&\multirow{2}{*}{$J^{P}$}&\multicolumn{4}{c}{Scheme~I}&\multicolumn{4}{c}{Scheme~II}\\
		\cmidrule(lr){3-6}
		\cmidrule(lr){7-10}
		&
		&Mass&$B_{s}^{*}B_{s}^{*}$&$B_{s}B_{s}^{*}$&$B_{s}B_{s}$
		&Mass&$B_{s}^{*}B_{s}^{*}$&$B_{s}B_{s}^{*}$&$B_{s}B_{s}$\\
		\midrule[1pt]
		%%%
		$ss\bar{b}\bar{b}$&$0^{+}$
		&$10697.1$
		&$\times$&&$\times$
		&$10792.1$
		&$\times$&&$167.6$\\
		&
		&$10928.8$
		&$410.8$&&$93.8$
		&$11154.3$
		&$725.3$&&$178.5$\\
		%%%
		&$1^{+}$
		&$10718.2$
		&&$\times$&
		&$10809.8$
		&&$64.3$\\
		%%%
		&$2^{+}$
		&$10742.5$
		&$\times$&&
		&$10834.1$
		&$44.4$\\
		%%%
		\bottomrule[1pt]
		\bottomrule[1pt]
	\end{tabular}
\end{table*}
%
%%%
%%% R:ssbb:I+II
%%%
\begin{table*}%[htbp]
	\centering
	\caption{The partial width ratios for the $ss\bar{b}\bar{b}$ tetraquarks. For each state, we choose one mode as the reference channel, and the partial width ratios of the other channels are calculated relative to this channel. The masses are all in unit of MeV.}
	\label{table:R:ssbb:I+II}
	\begin{tabular}{cccccccccc}
		\toprule[1pt]
		\toprule[1pt]
		\multirow{2}{*}{System}&\multirow{2}{*}{$J^{P}$}&\multicolumn{4}{c}{Scheme~I}&\multicolumn{4}{c}{Scheme~II}\\
		\cmidrule(lr){3-6}
		\cmidrule(lr){7-10}
		&
		&Mass&$B_{s}^{*}B_{s}^{*}$&$B_{s}B_{s}^{*}$&$B_{s}B_{s}$
		&Mass&$B_{s}^{*}B_{s}^{*}$&$B_{s}B_{s}^{*}$&$B_{s}B_{s}$\\
		\midrule[1pt]
		%%%
		$ss\bar{b}\bar{b}$&$0^{+}$
		&$10697.1$
		&$\times$&&$\times$
		&$10792.1$
		&$\times$&&$1$\\
		&
		&$10928.8$
		&$4.4$&&$1$
		&$11154.3$
		&$4.1$&&$1$\\
		%%%
		&$1^{+}$
		&$10718.2$
		&&$\times$&
		&$10809.8$
		&&$1$\\
		%%%
		&$2^{+}$
		&$10742.5$
		&$\times$&&
		&$10834.1$
		&$1$\\
		%%%
		\bottomrule[1pt]
		\bottomrule[1pt]
	\end{tabular}
\end{table*}
%
%%%
%%% eigenvector:sscb:I+II
%%%
\begin{table*}%[htbp]
	\centering
	\caption{The eigenvectors of the $ss\bar{c}\bar{b}$ tetraquark states in the $s\bar{c}{\otimes}s\bar{b}$ configuration. The masses are all in units of MeV.}
	\label{table:eigenvector:sscb:I+II}
	\begin{tabular}{cccccccccccccc}
		\toprule[1pt]
		\toprule[1pt]
		\multirow{2}{*}{System}&\multirow{2}{*}{$J^{P}$}&\multicolumn{5}{c}{Scheme~I}&\multicolumn{5}{c}{Scheme~II}\\
		\cmidrule(lr){3-7}
		\cmidrule(lr){8-12}
		&
		&Mass&$\bar{D}_{s}^{*}B_{s}^{*}$&$\bar{D}_{s}^{*}B_{s}$&$\bar{D}_{s}B_{s}^{*}$&$\bar{D}_{s}B_{s}$
		&Mass&$\bar{D}_{s}^{*}B_{s}^{*}$&$\bar{D}_{s}^{*}B_{s}$&$\bar{D}_{s}B_{s}^{*}$&$\bar{D}_{s}B_{s}$\\
		\midrule[1pt]
		%%%
		$ss\bar{c}\bar{b}$&$0^{+}$
		&$7404.4$&$0.145$&&&$0.642$
		&$7531.3$&$-0.043$&&&$0.606$\\
		&
		&$7597.3$&$0.750$&&&$0.068$
		&$7818.8$&$0.763$&&&$0.224$\\
		%%%
		&$1^{+}$
		&$7431.8$&$-0.049$&$0.179$&$-0.629$&
		&$7553.6$&$0.133$&$0.040$&$-0.581$\\
		&
		&$7503.3$&$0.191$&$0.536$&$0.110$&
		&$7603.5$&$0.249$&$0.515$&$0.085$\\
		&
		&$7569.4$&$0.679$&$-0.311$&$0.097$&
		&$7795.3$&$0.648$&$-0.388$&$0.268$\\
		%%%
		&$2^{+}$
		&$7534.3$&$0.577$&&&
		&$7633.8$&$0.577$\\
		%%%
		\bottomrule[1pt]
		\bottomrule[1pt]
	\end{tabular}
\end{table*}
%
%%%
%%% kc_i^2:sscb:I+II
%%%
\begin{table*}%[htbp]
	\centering
	\caption{The values of $k\cdot|c_{i}|^2$ for the $ss\bar{c}\bar{b}$ tetraquarks (in unit of MeV).}
	\label{table:kc_i^2:sscb:I+II}
	\begin{tabular}{ccccccccccccc}
		\toprule[1pt]
		\toprule[1pt]
		\multirow{2}{*}{System}&\multirow{2}{*}{$J^{P}$}&\multicolumn{5}{c}{Scheme~I}&\multicolumn{5}{c}{Scheme~II}\\
		\cmidrule(lr){3-7}
		\cmidrule(lr){8-12}
		&
		&Mass&$\bar{D}_{s}^{*}B_{s}^{*}$&$\bar{D}_{s}^{*}B_{s}$&$\bar{D}_{s}B_{s}^{*}$&$\bar{D}_{s}B_{s}$
		&Mass&$\bar{D}_{s}^{*}B_{s}^{*}$&$\bar{D}_{s}^{*}B_{s}$&$\bar{D}_{s}B_{s}^{*}$&$\bar{D}_{s}B_{s}$\\
		\midrule[1pt]
		%%%
		$ss\bar{c}\bar{b}$&$0^{+}$
		&$7404.4$
		&$\times$&&&$184.9$
		&$7531.3$
		&$0.2$&&&$279.4$\\
		&
		&$7597.3$
		&$260.2$&&&$4.1$
		&$7818.8$
		&$557.2$&&&$61.0$\\
		%%%
		&$1^{+}$
		&$7431.8$
		&$\times$&$\times$&$147.8$&
		&$7553.6$
		&$5.0$&$0.8$&$239.1$\\
		&
		&$7503.3$
		&$\times$&$78.3$&$7.2$&
		&$7603.5$
		&$29.9$&$164.1$&$5.9$\\
		&
		&$7569.4$
		&$165.0$&$51.1$&$6.9$&
		&$7795.3$
		&$385.6$&$150.0$&$80.7$\\
		%%%
		&$2^{+}$
		&$7534.3$
		&$47.8$&&&
		&$7633.8$
		&$190.8$\\
		%%%
		\bottomrule[1pt]
		\bottomrule[1pt]
	\end{tabular}
\end{table*}
%
%%%
%%% R:sscb:I+II
%%%
\begin{table*}%[htbp]
	\centering
	\caption{The partial width ratios for the $ss\bar{c}\bar{b}$ tetraquarks. For each state, we choose one mode as the reference channel, and the partial width ratios of the other channels are calculated relative to this channel. The masses are all in unit of MeV.}
	\label{table:R:sscb:I+II}
	\begin{tabular}{ccccccccccccccc}
		\toprule[1pt]
		\toprule[1pt]
		\multirow{2}{*}{System}&\multirow{2}{*}{$J^{P}$}&\multicolumn{5}{c}{Scheme~I}&\multicolumn{5}{c}{Scheme~II}\\
		\cmidrule(lr){3-7}
		\cmidrule(lr){8-12}
		&
		&Mass&$\bar{D}_{s}^{*}B_{s}^{*}$&$\bar{D}_{s}^{*}B_{s}$&$\bar{D}_{s}B_{s}^{*}$&$\bar{D}_{s}B_{s}$
		&Mass&$\bar{D}_{s}^{*}B_{s}^{*}$&$\bar{D}_{s}^{*}B_{s}$&$\bar{D}_{s}B_{s}^{*}$&$\bar{D}_{s}B_{s}$\\
		\midrule[1pt]
		%%%
		$ss\bar{c}\bar{b}$&$0^{+}$
		&$7404.4$
		&$\times$&&&$1$
		&$7531.3$
		&$0.0007$&&&$1$\\
		&
		&$7597.3$
		&$63.2$&&&$1$
		&$7818.8$
		&$9.1$&&&$1$\\
		%%%
		&$1^{+}$
		&$7431.8$
		&$\times$&$\times$&$1$&
		&$7553.6$
		&$0.02$&$0.003$&$1$\\
		&
		&$7503.3$
		&$\times$&$10.9$&$1$&
		&$7603.5$
		&$5.1$&$27.8$&$1$\\
		&
		&$7569.4$
		&$23.7$&$7.4$&$1$&
		&$7795.3$
		&$4.8$&$1.9$&$1$\\
		%%%
		&$2^{+}$
		&$7534.3$
		&$1$&&&
		&$7633.8$
		&$1$\\
		%%%
		\bottomrule[1pt]
		\bottomrule[1pt]
	\end{tabular}
\end{table*}
%
%%%
%%% mass:sscc+ssbb+:P1+P2
%%%
\begin{figure*}%[!h]
	\begin{tabular}{ccc}
		\includegraphics[width=450pt]{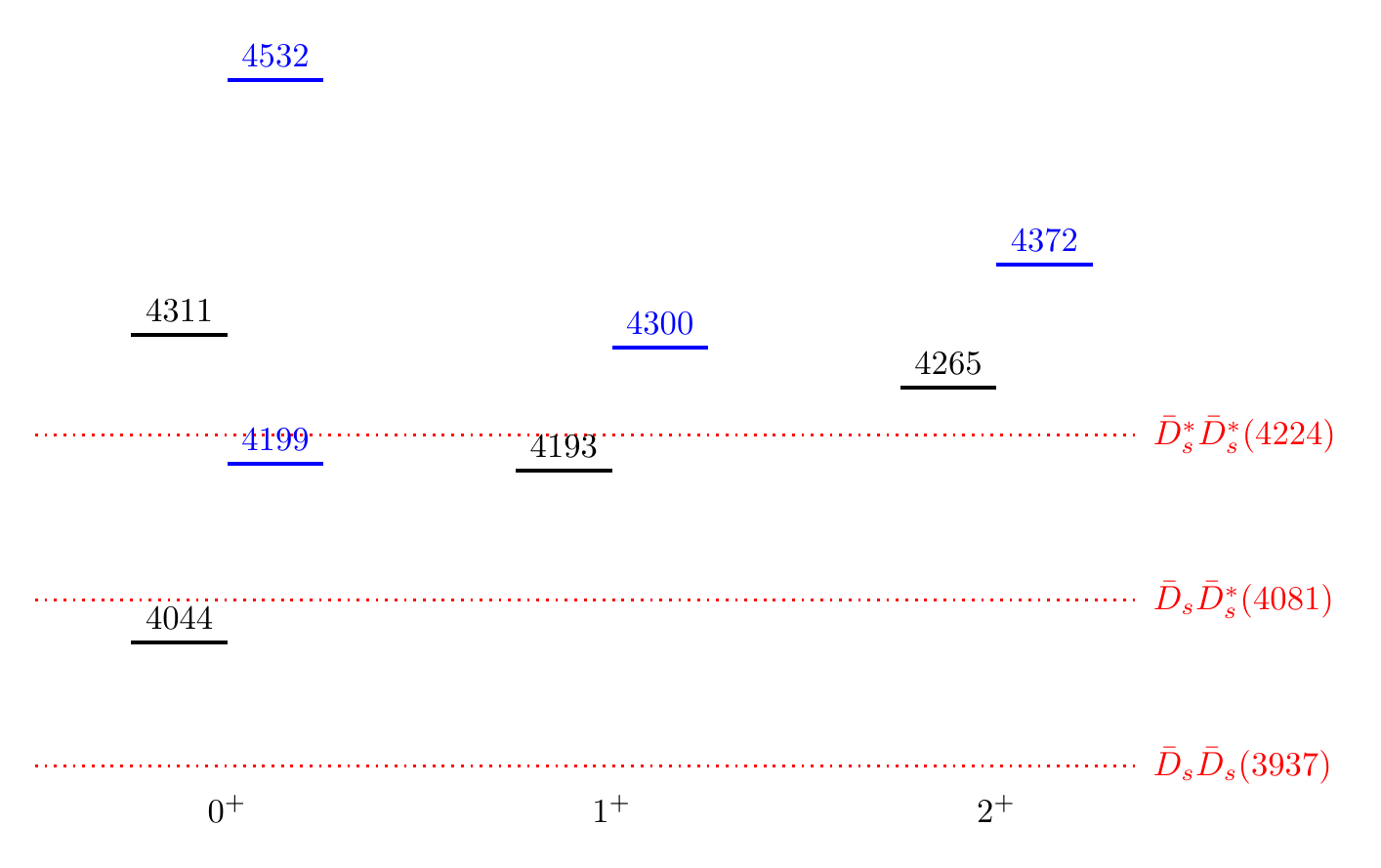}\\
		(a) $ss\bar{c}\bar{c}$ states\\
		&&\\
		\includegraphics[width=450pt]{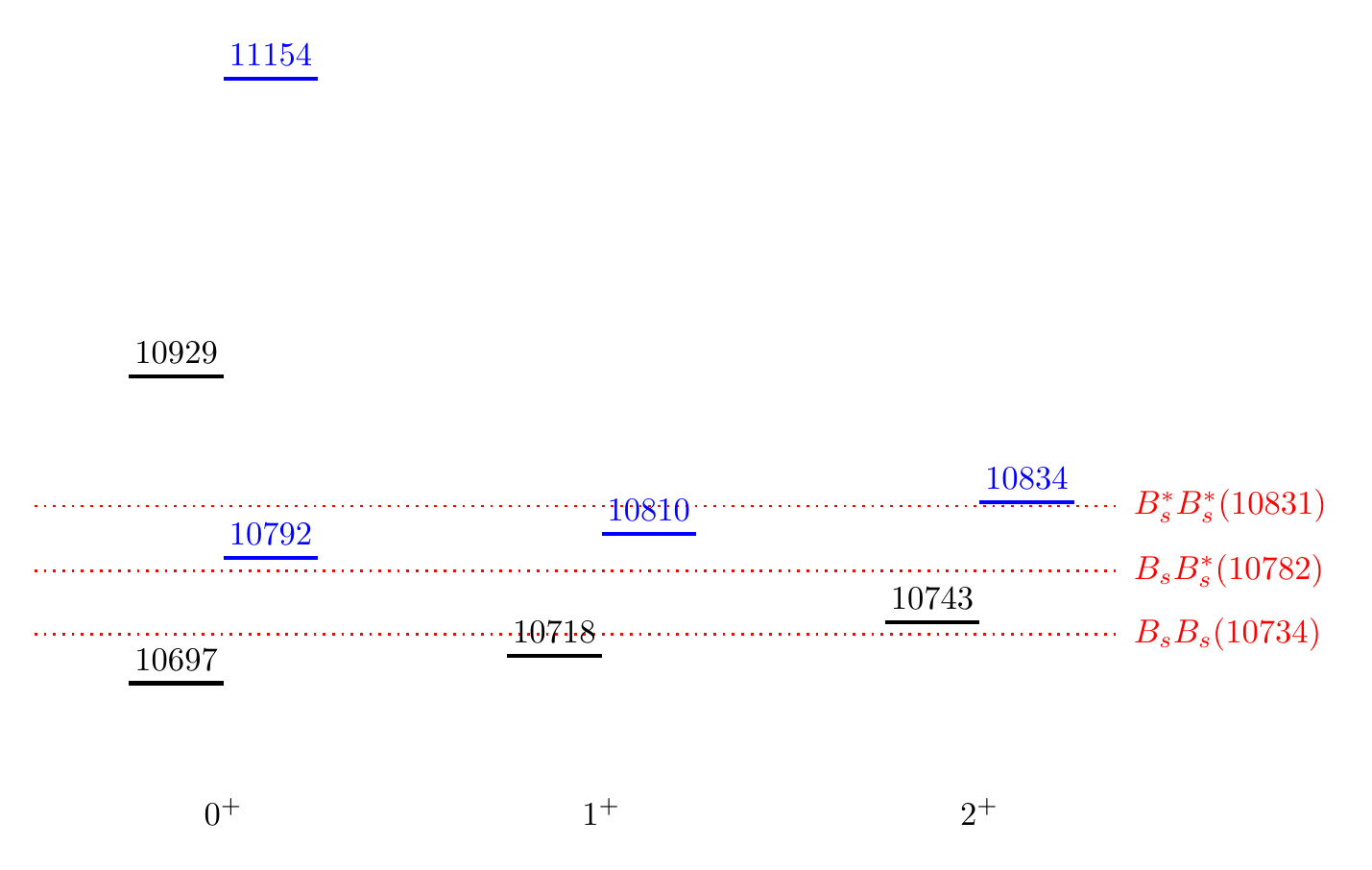}\\
		(b) $ss\bar{b}\bar{b}$ states\\
	\end{tabular}
	\caption{Mass spectra of the $ss\bar{c}\bar{c}$ and $ss\bar{b}\bar{b}$ tetraquark states in scheme~I (black) and scheme~II (blue). The dotted lines indicate various meson-meson thresholds. The masses are all in units of MeV.}
	\label{fig:sscc+ssbb}
\end{figure*}
%
%%%
%%% mass:sscb+:P1+P2
%%%
\begin{figure*}%[!h]
	\begin{tabular}{ccc}
		\includegraphics[width=450pt]{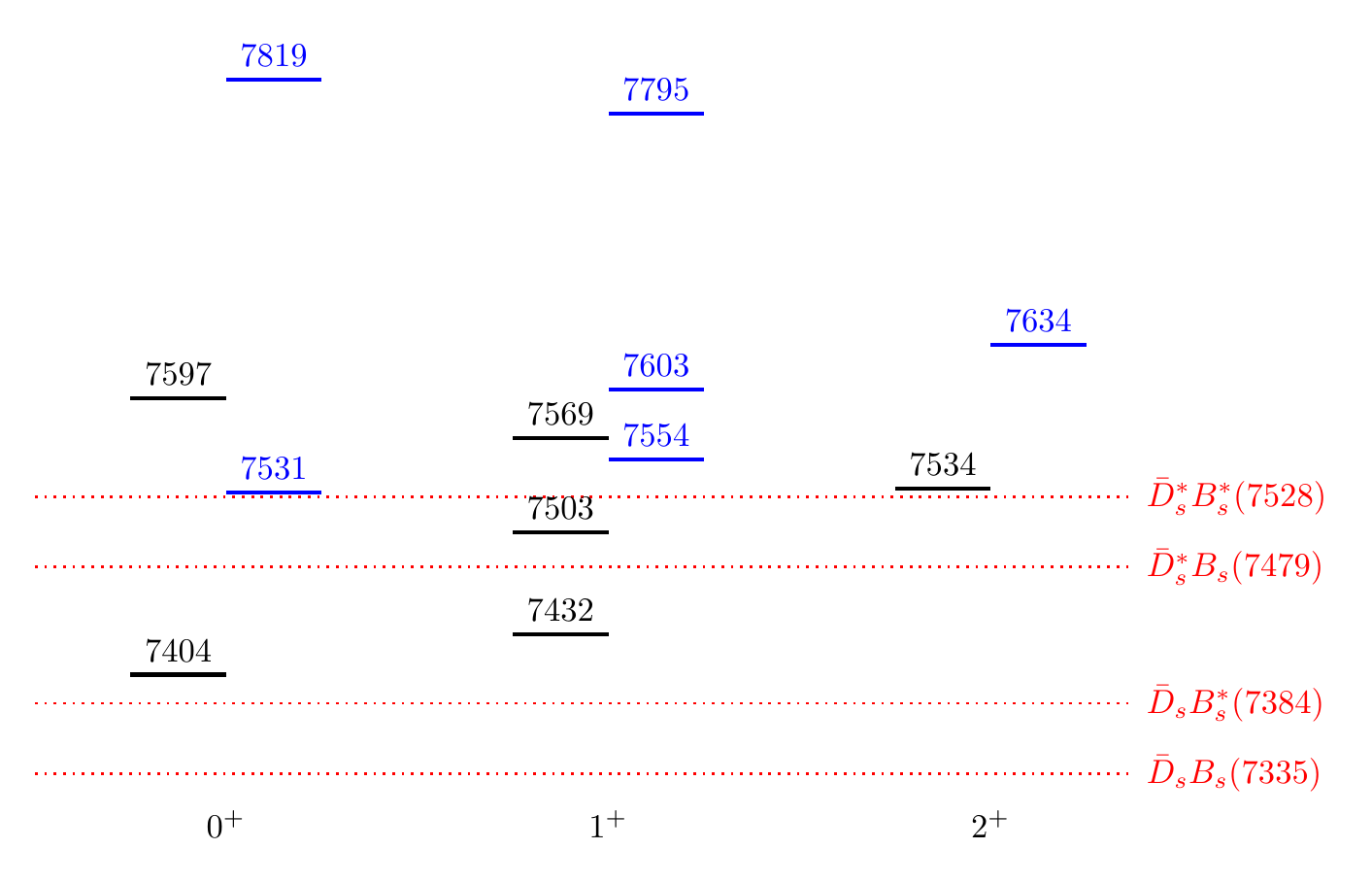}\\
	\end{tabular}
	\caption{Mass spectra of the $ss\bar{c}\bar{b}$ tetraquark states in scheme~I (black) and scheme~II (blue). The dotted lines indicate various meson-meson thresholds. The masses are all in units of MeV.}
	\label{fig:sscb}
\end{figure*}

We list the numerical results of the $ss\bar{Q}\bar{Q}$ in Table~\ref{table:mass:sscc+ssbb+sscb:I+II}--\ref{table:R:sscb:I+II}.
We also plot the relative position and possible decay channels in Figs.~\ref{fig:sscc+ssbb}--\ref{fig:sscb}.
The pattern of the mass spectrum is very similar to that of the $nn\bar{Q}\bar{Q}$ tetraquarks with isospin $I=1$.

First we focus on the $ss\bar{c}\bar{c}$ tetraquarks.
The ground state is $T_{I}(ss\bar{c}\bar{c},4043.7,0^{+})$.
%
%This state is dominated by color-triplet configuration ($57.8\%$)
%
It can decay into $\bar{D}_{s}\bar{D}_{s}$ in $S$-wave, and thus might be a wide state.
The most heavy state $T_{I}(ss\bar{c}\bar{c},4311.1,0^{+})$ lies above the $\bar{D}_{s}^{*}\bar{D}_{s}^{*}$ threshold.
It decays into $\bar{D}_{s}\bar{D}_{s}$ and $\bar{D}_{s}^{*}\bar{D}_{s}^{*}$ modes with the ratios
\begin{equation}
\frac{\Gamma[T_{I}(ss\bar{c}\bar{c},4311.1,0^{+}){\rightarrow}\bar{D}_{s}\bar{D}_{s}]}{\Gamma[T_{I}(ss\bar{c}\bar{c},4311.1,0^{+}){\rightarrow}\bar{D}_{s}^{*}\bar{D}_{s}^{*}]}
\sim0.0008\,.
\end{equation}
Thus the $\bar{D}_{s}^{*}\bar{D}_{s}^{*}$ mode is dominant.
The other two state $T_{I}(ss\bar{c}\bar{c},4192.6,1^{+})$ and $T_{I}(ss\bar{c}\bar{c},4264.5,2^{+})$ can decay into $\bar{D}_{s}\bar{D}_{s}^{*}$ and $\bar{D}_{s}^{*}\bar{D}_{s}^{*}$ modes respectively.

Next we turn to the $ss\bar{b}\bar{b}$ tetraquarks.
In scheme~I, the $T_{I}(ss\bar{b}\bar{b},10697.1,0^{+})$ and $T_{I}(ss\bar{b}\bar{b},10718.2,1^{+})$ lie below the $B_{s}B_{s}$ threshold, and the $T_{I}(ss\bar{b}\bar{b},10742.5,2^{+})$ lies just above the $B_{s}B_{s}$ threshold, which suggests that they are stable states.
However, they become heavier than their $S$-wave decay channels in scheme~II.
A detailed study with dynamical model, or experimental researches, is required to distinguish which of the two schemes gives better description of the $ss\bar{b}\bar{b}$ tetraquarks.
In both schemes, the three states are dominated by $\bar{3}_{c}\otimes3_{c}$ configuration.
Actually, their wave functions are nearly the same, except that they have different total spin, which is the reason for their different masses.
The highest state can decay into $B_{s}B_{s}$ and $B_{s}^{*}B_{s}^{*}$ modes, with nearly identical partial width ratios
\begin{equation}
\frac{\Gamma[T_{I}(ss\bar{b}\bar{b},10928.8,0^{+}){\rightarrow}B_{s}^{*}B_{s}^{*}]}{\Gamma[T_{I}(ss\bar{b}\bar{b},10928.8,0^{+}){\rightarrow}B_{s}B_{s}]}
\sim4.4
\end{equation}
and
\begin{equation}
\frac{\Gamma[T_{II}(ss\bar{b}\bar{b},11154.3,0^{+}){\rightarrow}B_{s}^{*}B_{s}^{*}]}{\Gamma[T_{II}(ss\bar{b}\bar{b},11154.3,0^{+}){\rightarrow}B_{s}B_{s}]}
\sim4.1\,.
\end{equation}

From Fig.~\ref{fig:sscb}, we see that the $ss\bar{c}\bar{b}$ tetraquarks are all above the $S$-wave decay channels and are probably broad states.
Among them, the $T_{I}(ss\bar{c}\bar{b},7534.3,2^{+})$ is slightly above the $\bar{D}_{s}^{*}B_{s}^{*}$.
Its decay width may be relatively narrow.
We also calculate the partial decay width ratios of the $ss\bar{c}\bar{b}$ tetraquarks.
It is interesting that some of the ratios are different in the two schemes.
For example, in scheme~I
\begin{equation}
\frac{\Gamma[T_{I}(ss\bar{c}\bar{b},7597.3,0^{+}){\rightarrow}\bar{D}_{s}^{*}B_{s}^{*}]}{\Gamma[T_{I}(ss\bar{c}\bar{b},7597.3,0^{+}){\rightarrow}\bar{D}_{s}B_{s}]}
\sim63.2
\end{equation}
and in scheme~II
\begin{equation}
\frac{\Gamma[T_{I}(ss\bar{c}\bar{b},7818.8,0^{+}){\rightarrow}\bar{D}_{s}^{*}B_{s}^{*}]}{\Gamma[T_{I}(ss\bar{c}\bar{b},7818.8,0^{+}){\rightarrow}\bar{D}_{s}B_{s}]}
\sim9.1\,,
\end{equation}
which can be used to distinguish the two schemes.
%

%%%%%%%%%%%%%%%%%%%%%%%%%%%%%%%%%%%%%%%%%%%%%%%%%%%%%%%%%%%%%%%%%%%%
\subsection{The $ns\bar{Q}\bar{Q}$ systems}
\label{sec:nsQQ}
%%%%%%%%%%%%%%%%%%%%%%%%%%%%%%%%%%%%%%%%%%%%%%%%%%%%%%%%%%%%%%%%%%%%

%%%
%%% mass:nscc+nsbb+nscb:I+II
%%%
\begin{table*}%[htbp]
	\centering
	\caption{Masses and eigenvectors of the $ns\bar{c}\bar{c}$, $ns\bar{b}\bar{b}$ and $ns\bar{c}\bar{b}$ tetraquarks. The masses are all in units of MeV.}
	\label{table:mass:nscc+nsbb+nscb:I+II}
	\begin{tabular}{ccccccc}
		\toprule[1pt]
		\toprule[1pt]
		\multirow{2}{*}{System}&\multirow{2}{*}{$J^{P}$}&\multicolumn{2}{c}{Scheme~I}&\multicolumn{2}{c}{Scheme~II}\\
		\cmidrule(lr){3-4}
		\cmidrule(lr){5-6}
		&&Mass&Eigenvector&Mass&Eigenvector\\
		\midrule[1pt]
		%%%
		$ns\bar{c}\bar{c}$&$0^{+}$
		&$3937.6$&$\{0.606,0.795\}$
		&$4085.7$&$\{0.410,0.912\}$\\
		&
		&$4209.3$&$\{0.795,-0.606\}$
		&$4436.2$&$\{0.912,-0.410\}$\\
		%%%
		&$1^{+}$
		&$3919.0$&$\{-0.534,-0.006,0.845\}$
		&$4051.5$&$\{0.284,0.005,-0.959\}$\\
		&
		&$4073.0$&$\{-0.032,-0.999,-0.027\}$
		&$4180.2$&$\{0.003,0.99997,0.007\}$\\
		&
		&$4086.5$&$\{0.845,-0.041,0.534\}$
		&$4328.9$&$\{0.959,-0.005,0.284\}$\\
		%%%
		&$2^{+}$
		&$4144.3$&$\{1\}$
		&$4251.5$&$\{1\}$\\
		%%%
		\midrule[1pt]
		%%%
		$ns\bar{b}\bar{b}$&$0^{+}$
		&$10586.4$&$\{0.163,0.987\}$
		&$10684.1$&$\{0.107,0.994\}$\\
		&
		&$10854.6$&$\{0.987,-0.163\}$
		&$11090.2$&$\{0.994,-0.107\}$\\
		%%%
		&$1^{+}$
		&$10473.1$&$\{0.082,0.005,-0.997\}$
		&$10569.0$&$\{0.056,0.005,-0.998\}$\\
		&
		&$10605.3$&$\{0.007,0.99996,0.006\}$
		&$10700.5$&$\{0.004,0.99998,0.005\}$\\
		&
		&$10778.7$&$\{0.997,-0.007,0.082\}$
		&$11016.1$&$\{0.998,-0.004,0.056\}$\\
		%%%
		&$2^{+}$
		&$10628.7$&$\{1\}$
		&$10723.9$&$\{1\}$\\
		%%%
		\midrule[1pt]
		%%%
		$ns\bar{c}\bar{b}$&$0^{+}$
		&$7156.5$&$\{0.647,0.029,0.038,0.761\}$
		&$7296.2$&$\{0.375,0.036,0.050,0.925\}$\\
		&
		&$7299.0$&$\{0.030,0.473,0.876,-0.087\}$
		&$7421.1$&$\{0.044,0.283,0.955,-0.080\}$\\
		&
		&$7333.2$&$\{-0.762,0.070,0.052,0.642\}$
		&$7547.8$&$\{-0.926,0.055,0.058,0.370\}$\\
		&
		&$7506.6$&$\{-0.023,-0.878,0.477,0.029\}$
		&$7738.5$&$\{-0.026,-0.957,0.287,0.032\}$\\
		%%%
		&$1^{+}$
		&$7212.8$&$\{0.423,-0.343,0.031,0.024,-0.025,0.838\}$
		&$7336.1$&$\{0.181,-0.185,0.035,0.023,-0.030,0.965\}$\\
		&
		&$7323.9$&$\{-0.210,0.057,-0.413,-0.589,0.635,0.180\}$
		&$7440.9$&$\{-0.044,0.030,-0.225,-0.675,0.698,0.060\}$\\
		&
		&$7330.5$&$\{-0.818,0.168,0.161,0.095,-0.225,0.466\}$
		&$7487.8$&$\{-0.089,-0.033,0.041,-0.720,-0.687,0.005\}$\\
		&
		&$7386.1$&$\{0.004,0.224,-0.056,0.749,0.615,0.089\}$
		&$7565.3$&$\{0.901,-0.351,-0.047,-0.090,-0.010,-0.233\}$\\
		&
		&$7416.2$&$\{0.329,0.894,0.062,-0.169,-0.144,0.198\}$
		&$7665.0$&$\{-0.381,-0.916,-0.042,0.039,0.048,-0.102\}$\\
		&
		&$7479.9$&$\{0.013,-0.040,0.892,-0.233,0.383,-0.038\}$
		&$7716.8$&$\{-0.014,0.042,-0.971,0.129,-0.194,0.037\}$\\
		%%%
		&$2^{+}$
		&$7415.1$&$\{0.297,0.955\}$
		&$7517.7$&$\{0.064,0.998\}$\\
		&
		&$7439.0$&$\{0.955,-0.297\}$
		&$7690.6$&$\{0.998,-0.064\}$\\
		%%%
		\bottomrule[1pt]
		\bottomrule[1pt]
	\end{tabular}
\end{table*}
%
%%%
%%% eigenvector:nscc:I+II
%%%
\begin{table*}%[htbp]
	\centering
	\caption{The eigenvectors of the $ns\bar{c}\bar{c}$ tetraquark states in the $n\bar{c}{\otimes}s\bar{c}$ configuration. The masses are all in units of MeV.}
	\label{table:eigenvector:nscc:I+II}
	\begin{tabular}{cccccccccccccc}
		\toprule[1pt]
		\toprule[1pt]
		\multirow{2}{*}{System}&\multirow{2}{*}{$J^{P}$}&\multicolumn{5}{c}{Scheme~I}&\multicolumn{5}{c}{Scheme~II}\\
		\cmidrule(lr){3-7}
		\cmidrule(lr){8-12}
		&
		&Mass&$\bar{D}^{*}\bar{D}_{s}^{*}$&$\bar{D}^{*}\bar{D}_{s}$&$\bar{D}\bar{D}_{s}^{*}$&$\bar{D}\bar{D}_{s}$
		&Mass&$\bar{D}^{*}\bar{D}_{s}^{*}$&$\bar{D}^{*}\bar{D}_{s}$&$\bar{D}\bar{D}_{s}^{*}$&$\bar{D}\bar{D}_{s}$\\
		\midrule[1pt]
		%%%
		$ns\bar{c}\bar{c}$&$0^{+}$
		&$3937.6$&$0.199$&&&$0.645$
		&$4085.7$&$0.026$&&&$0.623$\\
		&
		&$4209.3$&$0.737$&&&$0.022$
		&$4436.2$&$0.763$&&&$0.168$\\
		%%%
		&$1^{+}$
		&$3919.0$&$0.036$&$-0.464$&$0.460$&
		&$4051.5$&$-0.227$&$0.395$&$-0.391$\\
		&
		&$4073.0$&$-0.029$&$-0.413$&$-0.403$&
		&$4180.2$&$-0.005$&$-0.408$&$-0.409$\\
		&
		&$4086.5$&$0.706$&$0.174$&$-0.207$&
		&$4328.9$&$0.670$&$0.307$&$-0.311$\\
		%%%
		&$2^{+}$
		&$4144.3$&$0.577$&&&
		&$4251.5$&$0.577$\\
		%%%
		\bottomrule[1pt]
		\bottomrule[1pt]
	\end{tabular}
\end{table*}
%
%%%
%%% kc_i^2:nscc:I+II
%%%
\begin{table*}%[htbp]
	\centering
	\caption{The values of $k\cdot|c_{i}|^2$ for the $ns\bar{c}\bar{c}$ tetraquarks (in unit of MeV).}
	\label{table:kc_i^2:nscc:I+II}
	\begin{tabular}{ccccccccccccccc}
		\toprule[1pt]
		\toprule[1pt]
		\multirow{2}{*}{System}&\multirow{2}{*}{$J^{P}$}&\multicolumn{5}{c}{Scheme~I}&\multicolumn{5}{c}{Scheme~II}\\
		\cmidrule(lr){3-7}
		\cmidrule(lr){8-12}
		&
		&Mass&$\bar{D}^{*}\bar{D}_{s}^{*}$&$\bar{D}^{*}\bar{D}_{s}$&$\bar{D}\bar{D}_{s}^{*}$&$\bar{D}\bar{D}_{s}$
		&Mass&$\bar{D}^{*}\bar{D}_{s}^{*}$&$\bar{D}^{*}\bar{D}_{s}$&$\bar{D}\bar{D}_{s}^{*}$&$\bar{D}\bar{D}_{s}$\\
		\midrule[1pt]
		%%%
		$ns\bar{c}\bar{c}$&$0^{+}$
		&$3937.6$
		&$\times$&&&$185.3$
		&$4085.7$
		&$\times$&&&$273.5$\\
		&
		&$4209.3$
		&$233.6$&&&$0.4$
		&$4436.2$
		&$478.6$&&&$31.3$\\
		%%%
		&$1^{+}$
		&$3919.0$
		&$\times$&$\times$&$\times$&
		&$4051.5$
		&$\times$&$60.4$&$58.0$\\
		&
		&$4073.0$
		&$\times$&$75.1$&$70.3$&
		&$4180.2$
		&$0.008$&$107.1$&$106.8$\\
		&
		&$4086.5$
		&$\times$&$14.2$&$20.0$&
		&$4328.9$
		&$297.3$&$80.7$&$82.5$\\
		%%%
		&$2^{+}$
		&$4144.3$
		&$73.7$&&&
		&$4251.5$
		&$174.4$\\
		%%%
		\bottomrule[1pt]
		\bottomrule[1pt]
	\end{tabular}
\end{table*}
%
%%%
%%% R:nscc:I+II
%%%
\begin{table*}%[htbp]
	\centering
	\caption{The partial width ratios for the $ns\bar{c}\bar{c}$ tetraquarks. For each state, we choose one mode as the reference channel, and the partial width ratios of the other channels are calculated relative to this channel. The masses are all in unit of MeV.}
	\label{table:R:nscc:I+II}
	\begin{tabular}{cccccccccccccc}
		\toprule[1pt]
		\toprule[1pt]
		\multirow{2}{*}{System}&\multirow{2}{*}{$J^{P}$}&\multicolumn{5}{c}{Scheme~I}&\multicolumn{5}{c}{Scheme~II}\\
		\cmidrule(lr){3-7}
		\cmidrule(lr){8-12}
		&
		&Mass&$\bar{D}^{*}\bar{D}_{s}^{*}$&$\bar{D}^{*}\bar{D}_{s}$&$\bar{D}\bar{D}_{s}^{*}$&$\bar{D}\bar{D}_{s}$
		&Mass&$\bar{D}^{*}\bar{D}_{s}^{*}$&$\bar{D}^{*}\bar{D}_{s}$&$\bar{D}\bar{D}_{s}^{*}$&$\bar{D}\bar{D}_{s}$\\
		\midrule[1pt]
		%%%
		$ns\bar{c}\bar{c}$&$0^{+}$
		&$3937.6$
		&$\times$&&&$1$
		&$4085.7$
		&$\times$&&&$1$\\
		&
		&$4209.3$
		&$569.1$&&&$1$
		&$4436.2$
		&$15.3$&&&$1$\\
		%%%
		&$1^{+}$
		&$3919.0$
		&$\times$&$\times$&$\times$&
		&$4051.5$
		&$\times$&$1.04$&$1$\\
		&
		&$4073.0$
		&$\times$&$1.1$&$1$&
		&$4180.2$
		&$0.00007$&$1.003$&$1$\\
		&
		&$4086.5$
		&$\times$&$0.7$&$1$&
		&$4328.9$
		&$3.6$&$0.98$&$1$\\
		%%%
		&$2^{+}$
		&$4144.3$
		&$1$&&&
		&$4251.5$
		&$1$\\
		%%%
		\bottomrule[1pt]
		\bottomrule[1pt]
	\end{tabular}
\end{table*}
%
%%%
%%% eigenvector:nsbb:I+II
%%%
\begin{table*}%[htbp]
	\centering
	\caption{The eigenvectors of the $ns\bar{b}\bar{b}$ tetraquark states in the $n\bar{b}{\otimes}s\bar{b}$ configuration. The masses are all in units of MeV.}
	\label{table:eigenvector:nsbb:I+II}
	\begin{tabular}{cccccccccccccc}
		\toprule[1pt]
		\toprule[1pt]
		\multirow{2}{*}{System}&\multirow{2}{*}{$J^{P}$}&\multicolumn{5}{c}{Scheme~I}&\multicolumn{5}{c}{Scheme~II}\\
		\cmidrule(lr){3-7}
		\cmidrule(lr){8-12}
		&
		&Mass&$B^{*}B_{s}^{*}$&$B^{*}B_{s}$&$BB_{s}^{*}$&$BB_{s}$
		&Mass&$B^{*}B_{s}^{*}$&$B^{*}B_{s}$&$BB_{s}^{*}$&$BB_{s}$\\
		\midrule[1pt]
		%%%
		$ns\bar{b}\bar{b}$&$0^{+}$
		&$10586.4$&$-0.170$&&&$0.560$
		&$10684.1$&$-0.212$&&&$0.541$\\
		&
		&$10854.6$&$0.745$&&&$0.321$
		&$11090.2$&$0.734$&&&$0.353$\\
		%%%
		&$1^{+}$&
		$10473.1$&$-0.360$&$0.323$&$-0.319$&
		&$10569.0$&$-0.375$&$0.313$&$-0.309$\\
		&
		&$10605.3$&$-0.006$&$-0.409$&$-0.407$&
		&$10700.5$&$-0.004$&$-0.408$&$-0.408$\\
		&
		&$10778.7$&$0.609$&$0.380$&$-0.386$&
		&$11016.1$&$0.599$&$0.390$&$-0.393$\\
		%%%
		&$2^{+}$
		&$10628.7$&$0.577$&&&
		&$10723.9$&$0.577$\\
		%%%
		\bottomrule[1pt]
		\bottomrule[1pt]
	\end{tabular}
\end{table*}
%
%%%
%%% kc_i^2:nsbb:I+II
%%%
\begin{table*}%[htbp]
	\centering
	\caption{The values of $k\cdot|c_{i}|^2$ for the $ns\bar{b}\bar{b}$ tetraquarks (in unit of MeV).}
	\label{table:kc_i^2:nsbb:I+II}
	\begin{tabular}{ccccccccccccc}
		\toprule[1pt]
		\toprule[1pt]
		\multirow{2}{*}{System}&\multirow{2}{*}{$J^{P}$}&\multicolumn{5}{c}{Scheme~I}&\multicolumn{5}{c}{Scheme~II}\\
		\cmidrule(lr){3-7}
		\cmidrule(lr){8-12}
		&
		&Mass&$B^{*}B_{s}^{*}$&$B^{*}B_{s}$&$BB_{s}^{*}$&$BB_{s}$
		&Mass&$B^{*}B_{s}^{*}$&$B^{*}B_{s}$&$BB_{s}^{*}$&$BB_{s}$\\
		\midrule[1pt]
		%%%
		$ns\bar{b}\bar{b}$&$0^{+}$
		&$10586.4$
		&$\times$&&&$\times$
		&$10684.1$
		&$\times$&&&$131.4$\\
		&
		&$10854.6$
		&$436.2$&&&$109.4$
		&$11090.2$
		&$744.5$&&&$193.1$\\
		%%%
		&$1^{+}$&
		$10473.1$
		&$\times$&$\times$&$\times$&
		&$10569.0$
		&$\times$&$\times$&$\times$\\
		&
		&$10605.3$
		&$\times$&$\times$&$\times$&
		&$10700.5$
		&$\times$&$36.6$&$28.9$\\
		&
		&$10778.7$
		&$168.9$&$99.0$&$100.0$&
		&$11016.1$
		&$439.9$&$201.8$&$204.1$\\
		%%%
		&$2^{+}$
		&$10628.7$
		&$\times$&&&
		&$10723.9$
		&$\times$\\
		%%%
		\bottomrule[1pt]
		\bottomrule[1pt]
	\end{tabular}
\end{table*}
%
%%%
%%% R:nsbb:I+II
%%%
\begin{table*}%[htbp]
	\centering
	\caption{The partial width ratios for the $ns\bar{b}\bar{b}$ tetraquarks. For each state, we choose one mode as the reference channel, and the partial width ratios of the other channels are calculated relative to this channel. The masses are all in unit of MeV.}
	\label{table:R:nsbb:I+II}
	\begin{tabular}{ccccccccccccc}
		\toprule[1pt]
		\toprule[1pt]
		\multirow{2}{*}{System}&\multirow{2}{*}{$J^{P}$}&\multicolumn{5}{c}{Scheme~I}&\multicolumn{5}{c}{Scheme~II}\\
		\cmidrule(lr){3-7}
		\cmidrule(lr){8-12}
		&
		&Mass&$B^{*}B_{s}^{*}$&$B^{*}B_{s}$&$BB_{s}^{*}$&$BB_{s}$
		&Mass&$B^{*}B_{s}^{*}$&$B^{*}B_{s}$&$BB_{s}^{*}$&$BB_{s}$\\
		\midrule[1pt]
		%%%
		$ns\bar{b}\bar{b}$&$0^{+}$
		&$10586.4$
		&$\times$&&&$\times$
		&$10684.1$
		&$\times$&&&$1$\\
		&
		&$10854.6$
		&$4.0$&&&$1$
		&$11090.2$
		&$3.9$&&&$1$\\
		%%%
		&$1^{+}$&
		$10473.1$
		&$\times$&$\times$&$\times$&
		&$10569.0$
		&$\times$&$\times$&$\times$\\
		&
		&$10605.3$
		&$\times$&$\times$&$\times$&
		&$10700.5$
		&$\times$&$1.3$&$1$\\
		&
		&$10778.7$
		&$1.7$&$1.0$&$1$&
		&$11016.1$
		&$2.2$&$1.0$&$1$\\
		%%%
		&$2^{+}$
		&$10628.7$
		&$\times$&&&
		&$10723.9$
		&$\times$\\
		%%%
		\bottomrule[1pt]
		\bottomrule[1pt]
	\end{tabular}
\end{table*}
%
%%%
%%% eigenvector:nscb:n1c3xs2b4:I+II
%%%
\begin{table*}%[htbp]
	\centering
	\caption{The eigenvectors of the $ns\bar{c}\bar{b}$ tetraquark states in the $n\bar{c}{\otimes}s\bar{b}$ configuration. The masses are all in units of MeV.}
	\label{table:eigenvector:nscb:n1c3xs2b4:I+II}
	\begin{tabular}{cccccccccccccc}
		\toprule[1pt]
		\toprule[1pt]
		\multirow{2}{*}{System}&\multirow{2}{*}{$J^{P}$}&\multicolumn{5}{c}{Scheme~I}&\multicolumn{5}{c}{Scheme~II}\\
		\cmidrule(lr){3-7}
		\cmidrule(lr){8-12}
		&
		&Mass&$\bar{D}^{*}B_{s}^{*}$&$\bar{D}^{*}B_{s}$&$\bar{D}B_{s}^{*}$&$\bar{D}B_{s}$
		&Mass&$\bar{D}^{*}B_{s}^{*}$&$\bar{D}^{*}B_{s}$&$\bar{D}B_{s}^{*}$&$\bar{D}B_{s}$\\
		\midrule[1pt]
		%%%
		$ns\bar{c}\bar{b}$&$0^{+}$
		&$7156.5$
		&$0.126$&&&$0.708$
		&$7296.2$
		&$-0.320$&&&$-0.572$\\
		&
		&$7299.0$
		&$-0.026$&&&$-0.627$
		&$7421.1$
		&$0.134$&&&$-0.601$\\
		&
		&$7333.2$
		&$-0.666$&&&$0.299$
		&$7547.8$
		&$-0.585$&&&$0.496$\\
		&
		&$7506.6$
		&$-0.735$&&&$-0.128$
		&$7738.5$
		&$-0.733$&&&$-0.256$\\
		%%%
		&$1^{+}$
		&$7212.8$
		&$0.152$&$-0.148$&$0.655$&
		&$7336.1$
		&$0.295$&$-0.263$&$0.491$\\
		&
		&$7323.9$
		&$0.127$&$-0.039$&$-0.685$&
		&$7440.9$
		&$0.197$&$-0.012$&$-0.589$\\
		&
		&$7330.5$
		&$0.289$&$-0.630$&$-0.237$&
		&$7487.8$
		&$-0.273$&$-0.575$&$-0.115$\\
		&
		&$7386.1$
		&$0.384$&$0.574$&$0.042$&
		&$7565.3$
		&$-0.329$&$0.424$&$0.544$\\
		&
		&$7416.2$
		&$0.574$&$0.362$&$-0.120$&
		&$7665.0$
		&$-0.575$&$-0.517$&$0.109$\\
		&
		&$7479.9$
		&$0.633$&$-0.346$&$0.171$&
		&$7716.8$
		&$-0.600$&$0.391$&$-0.302$\\
		%%%
		&$2^{+}$
		&$7415.1$
		&$0.794$&&&
		&$7517.7$
		&$0.629$\\
		&
		&$7439.0$
		&$0.608$&&&
		&$7690.6$
		&$0.778$\\
		%%%
		\bottomrule[1pt]
		\bottomrule[1pt]
	\end{tabular}
\end{table*}
%
%%%
%%% eigenvector:nscb:n1b4xs2c3:I+II
%%%
\begin{table*}%[htbp]
	\centering
	\caption{The eigenvectors of the $ns\bar{c}\bar{b}$ tetraquark states in the $n\bar{b}{\otimes}s\bar{c}$ configuration. The masses are all in units of MeV.}
	\label{table:eigenvector:nscb:n1b4xs2c3:I+II}
	\begin{tabular}{cccccccccccccc}
		\toprule[1pt]
		\toprule[1pt]
		\multirow{2}{*}{System}&\multirow{2}{*}{$J^{P}$}&\multicolumn{5}{c}{Scheme~I}&\multicolumn{5}{c}{Scheme~II}\\
		\cmidrule(lr){3-7}
		\cmidrule(lr){8-12}
		&
		&Mass&$B^{*}\bar{D}_{s}^{*}$&$B^{*}\bar{D}_{s}$&$B\bar{D}_{s}^{*}$&$B\bar{D}_{s}$
		&Mass&$B^{*}\bar{D}_{s}^{*}$&$B^{*}\bar{D}_{s}$&$B\bar{D}_{s}^{*}$&$B\bar{D}_{s}$\\
		\midrule[1pt]
		%%%
		$ns\bar{c}\bar{b}$&$0^{+}$
		&$7156.5$
		&$0.107$&&&$0.646$
		&$7296.2$
		&$-0.298$&&&$-0.493$\\
		&
		&$7299.0$
		&$0.137$&&&$0.635$
		&$7421.1$
		&$-0.018$&&&$0.584$\\
		&
		&$7333.2$
		&$-0.598$&&&$0.407$
		&$7547.8$
		&$-0.541$&&&$0.599$\\
		&
		&$7506.6$
		&$0.783$&&&$0.112$
		&$7738.5$
		&$0.786$&&&$0.238$\\
		%%%
		&$1^{+}$
		&$7212.8$
		&$-0.136$&$0.596$&$-0.128$&
		&$7336.1$
		&$-0.279$&$0.426$&$-0.236$\\
		&
		&$7323.9$
		&$-0.086$&$0.500$&$-0.261$&
		&$7440.9$
		&$0.114$&$0.549$&$-0.048$\\
		&
		&$7330.5$
		&$-0.286$&$-0.576$&$-0.447$&
		&$7487.8$
		&$-0.240$&$0.042$&$0.443$\\
		&
		&$7386.1$
		&$0.054$&$-0.169$&$-0.438$&
		&$7565.3$
		&$0.266$&$0.649$&$0.465$\\
		&
		&$7416.2$
		&$-0.621$&$-0.116$&$0.634$&
		&$7665.0$
		&$0.566$&$0.140$&$-0.612$\\
		&
		&$7479.9$
		&$0.710$&$-0.146$&$0.351$&
		&$7716.8$
		&$-0.679$&$0.273$&$-0.395$\\
		%%%
		&$2^{+}$
		&$7415.1$
		&$-0.308$&&&
		&$7517.7$
		&$-0.524$\\
		&
		&$7439.0$
		&$0.951$&&&
		&$7690.6$
		&$0.852$\\
		%%%
		\bottomrule[1pt]
		\bottomrule[1pt]
	\end{tabular}
\end{table*}
%
%%%
%%% kc_i^2:nscb:I+II
%%%
\begin{table*}%[htbp]
	\centering
	\caption{The values of $k\cdot|c_{i}|^2$ for the $ns\bar{c}\bar{b}$ tetraquarks (in unit of MeV).}
	\label{table:kc_i^2:nscb:I+II}
	\begin{tabular}{ccccccccccccc}
		\toprule[1pt]
		\toprule[1pt]
		\multirow{2}{*}{System}&\multirow{2}{*}{$J^{P}$}&\multicolumn{5}{c}{Scheme~I}&\multicolumn{5}{c}{Scheme~II}\\
		\cmidrule(lr){3-7}
		\cmidrule(lr){8-12}
		&
		&Mass&$\bar{D}^{*}B_{s}^{*}$&$\bar{D}^{*}B_{s}$&$\bar{D}B_{s}^{*}$&$\bar{D}B_{s}$
		&Mass&$\bar{D}^{*}B_{s}^{*}$&$\bar{D}^{*}B_{s}$&$\bar{D}B_{s}^{*}$&$\bar{D}B_{s}$\\
		\midrule[1pt]
		%%%
		$ns\bar{c}\bar{b}$&$0^{+}$
		&$7156.5$
		&$\times$&&&$\times$
		&$7296.2$
		&$\times$&&&$136.2$\\
		&
		&$7299.0$
		&$\times$&&&$167.7$
		&$7421.1$
		&$\times$&&&$263.8$\\
		&
		&$7333.2$
		&$\times$&&&$47.1$
		&$7547.8$
		&$207.8$&&&$234.9$\\
		&
		&$7506.6$
		&$267.1$&&&$14.5$
		&$7738.5$
		&$527.0$&&&$80.5$\\
		%%%
		&$1^{+}$
		&$7212.8$
		&$\times$&$\times$&$\times$&
		&$7336.1$
		&$\times$&$\times$&$93.3$\\
		&
		&$7323.9$
		&$\times$&$\times$&$159.4$&
		&$7440.9$
		&$8.7$&$0.07$&$232.6$\\
		&
		&$7330.5$
		&$\times$&$\times$&$20.6$&
		&$7487.8$
		&$32.4$&$190.9$&$10.2$\\
		&
		&$7386.1$
		&$\times$&$58.5$&$1.0$&
		&$7565.3$
		&$70.5$&$135.6$&$267.3$\\
		&
		&$7416.2$
		&$\times$&$45.3$&$8.9$&
		&$7665.0$
		&$282.4$&$250.9$&$12.7$\\
		&
		&$7479.9$
		&$162.8$&$66.8$&$22.0$&
		&$7716.8$
		&$340.3$&$156.4$&$103.6$\\
		%%%
		&$2^{+}$
		&$7415.1$
		&$\times$&&&
		&$7517.7$
		&$208.5$\\
		&
		&$7439.0$
		&$77.7$&&&
		&$7690.6$
		&$544.2$\\
		%%%
		\midrule[1pt]
		\midrule[1pt]
		%%%
		\multirow{2}{*}{System}&\multirow{2}{*}{$J^{P}$}&\multicolumn{5}{c}{Scheme~I}&\multicolumn{5}{c}{Scheme~II}\\
		\cmidrule(lr){3-7}
		\cmidrule(lr){8-12}
		&
		&Mass&$B^{*}\bar{D}_{s}^{*}$&$B^{*}\bar{D}_{s}$&$B\bar{D}_{s}^{*}$&$B\bar{D}_{s}$
		&Mass&$B^{*}\bar{D}_{s}^{*}$&$B^{*}\bar{D}_{s}$&$B\bar{D}_{s}^{*}$&$B\bar{D}_{s}$\\
		\midrule[1pt]
		%%%
		$ns\bar{c}\bar{b}$&$0^{+}$
		&$7156.5$
		&$\times$&&&$\times$
		&$7296.2$
		&$\times$&&&$90.9$\\
		&
		&$7299.0$
		&$\times$&&&$155.3$
		&$7421.1$
		&$\times$&&&$243.7$\\
		&
		&$7333.2$
		&$\times$&&&$82.7$
		&$7547.8$
		&$170.9$&&&$339.5$\\
		&
		&$7506.6$
		&$282.8$&&&$11.0$
		&$7738.5$
		&$601.5$&&&$69.5$\\
		%%%
		&$1^{+}$
		&$7212.8$
		&$\times$&$\times$&$\times$&
		&$7336.1$
		&$\times$&$64.1$&$\times$\\
		&
		&$7323.9$
		&$\times$&$74.7$&$\times$&
		&$7440.9$
		&$1.4$&$198.4$&$0.9$\\
		&
		&$7330.5$
		&$\times$&$109.2$&$\times$&
		&$7487.8$
		&$22.6$&$1.4$&$106.4$\\
		&
		&$7386.1$
		&$\times$&$14.8$&$\times$&
		&$7565.3$
		&$44.6$&$379.9$&$158.0$\\
		&
		&$7416.2$
		&$\times$&$8.0$&$109.6$&
		&$7665.0$
		&$269.8$&$20.7$&$345.5$\\
		&
		&$7479.9$
		&$182.6$&$15.8$&$63.8$&
		&$7716.8$
		&$431.6$&$84.8$&$157.8$\\
		%%%
		&$2^{+}$
		&$7415.1$
		&$\times$&&&
		&$7517.7$
		&$136.4$\\
		&
		&$7439.0$
		&$75.0$&&&
		&$7690.6$
		&$646.2$\\
		%%%
		\bottomrule[1pt]
		\bottomrule[1pt]
	\end{tabular}
\end{table*}
%
%%%
%%% R:nscb:I+II
%%%
\begin{table*}%[htbp]
	\centering
	\caption{The partial width ratios for the $ns\bar{c}\bar{b}$ tetraquarks. For each state, we choose one mode as the reference channel, and the partial width ratios of the other channels are calculated relative to this channel. The masses are all in unit of MeV.}
	\label{table:R:nscb:I+II}
	\begin{tabular}{ccccccccccccccc}
		\toprule[1pt]
		\toprule[1pt]
		\multirow{2}{*}{System}&\multirow{2}{*}{$J^{P}$}&\multicolumn{5}{c}{Scheme~I}&\multicolumn{5}{c}{Scheme~II}\\
		\cmidrule(lr){3-7}
		\cmidrule(lr){8-12}
		&
		&Mass&$\bar{D}^{*}B_{s}^{*}$&$\bar{D}^{*}B_{s}$&$\bar{D}B_{s}^{*}$&$\bar{D}B_{s}$
		&Mass&$\bar{D}^{*}B_{s}^{*}$&$\bar{D}^{*}B_{s}$&$\bar{D}B_{s}^{*}$&$\bar{D}B_{s}$\\
		\midrule[1pt]
		%%%
		$ns\bar{c}\bar{b}$&$0^{+}$
		&$7156.5$
		&$\times$&&&$\times$
		&$7296.2$
		&$\times$&&&$1$\\
		&
		&$7299.0$
		&$\times$&&&$1$
		&$7421.1$
		&$\times$&&&$1$\\
		&
		&$7333.2$
		&$\times$&&&$1$
		&$7547.8$
		&$0.9$&&&$1$\\
		&
		&$7506.6$
		&$18.4$&&&$1$
		&$7738.5$
		&$6.6$&&&$1$\\
		%%%
		&$1^{+}$
		&$7212.8$
		&$\times$&$\times$&$\times$&
		&$7336.1$
		&$\times$&$\times$&$1$\\
		&
		&$7323.9$
		&$\times$&$\times$&$1$&
		&$7440.9$
		&$0.04$&$0.0003$&$1$\\
		&
		&$7330.5$
		&$\times$&$\times$&$1$&
		&$7487.8$
		&$3.2$&$18.8$&$1$\\
		&
		&$7386.1$
		&$\times$&$61.5$&$1$&
		&$7565.3$
		&$0.3$&$0.5$&$1$\\
		&
		&$7416.2$
		&$\times$&$5.1$&$1$&
		&$7665.0$
		&$22.3$&$19.8$&$1$\\
		&
		&$7479.9$
		&$7.4$&$3.0$&$1$&
		&$7716.8$
		&$3.3$&$1.5$&$1$\\
		%%%
		&$2^{+}$
		&$7415.1$
		&$\times$&&&
		&$7517.7$
		&$1$\\
		&
		&$7439.0$
		&$1$&&&
		&$7690.6$
		&$1$\\
		%%%
		\midrule[1pt]
		\midrule[1pt]
		%%%
		\multirow{2}{*}{System}&\multirow{2}{*}{$J^{P}$}&\multicolumn{5}{c}{Scheme~I}&\multicolumn{5}{c}{Scheme~II}\\
		\cmidrule(lr){3-7}
		\cmidrule(lr){8-12}
		&
		&Mass&$B^{*}\bar{D}_{s}^{*}$&$B^{*}\bar{D}_{s}$&$B\bar{D}_{s}^{*}$&$B\bar{D}_{s}$
		&Mass&$B^{*}\bar{D}_{s}^{*}$&$B^{*}\bar{D}_{s}$&$B\bar{D}_{s}^{*}$&$B\bar{D}_{s}$\\
		\midrule[1pt]
		%%%
		$ns\bar{c}\bar{b}$&$0^{+}$
		&$7156.5$
		&$\times$&&&$\times$
		&$7296.2$
		&$\times$&&&$1$\\
		&
		&$7299.0$
		&$\times$&&&$1$
		&$7421.1$
		&$\times$&&&$1$\\
		&
		&$7333.2$
		&$\times$&&&$1$
		&$7547.8$
		&$0.5$&&&$1$\\
		&
		&$7506.6$
		&$25.8$&&&$1$
		&$7738.5$
		&$8.7$&&&$1$\\
		%%%
		&$1^{+}$
		&$7212.8$
		&$\times$&$\times$&$\times$&
		&$7336.1$
		&$\times$&$1$&$\times$\\
		&
		&$7323.9$
		&$\times$&$1$&$\times$&
		&$7440.9$
		&$0.007$&$1$&$0.005$\\
		&
		&$7330.5$
		&$\times$&$1$&$\times$&
		&$7487.8$
		&$16.7$&$1$&$78.4$\\
		&
		&$7386.1$
		&$\times$&$1$&$\times$&
		&$7565.3$
		&$0.1$&$1$&$0.4$\\
		&
		&$7416.2$
		&$\times$&$1$&$13.6$&
		&$7665.0$
		&$13.0$&$1$&$16.7$\\
		&
		&$7479.9$
		&$11.5$&$1$&$4.0$&
		&$7716.8$
		&$5.1$&$1$&$1.9$\\
		%%%
		&$2^{+}$
		&$7415.1$
		&$\times$&&&
		&$7517.7$
		&$1$\\
		&
		&$7439.0$
		&$1$&&&
		&$7690.6$
		&$1$\\
		%%%
		\bottomrule[1pt]
		\bottomrule[1pt]
	\end{tabular}
\end{table*}
%
%%%
%%% mass:nscc+nsbb+:P1+P2
%%%
\begin{figure*}%[!h]
	\begin{tabular}{ccc}
		\includegraphics[width=450pt]{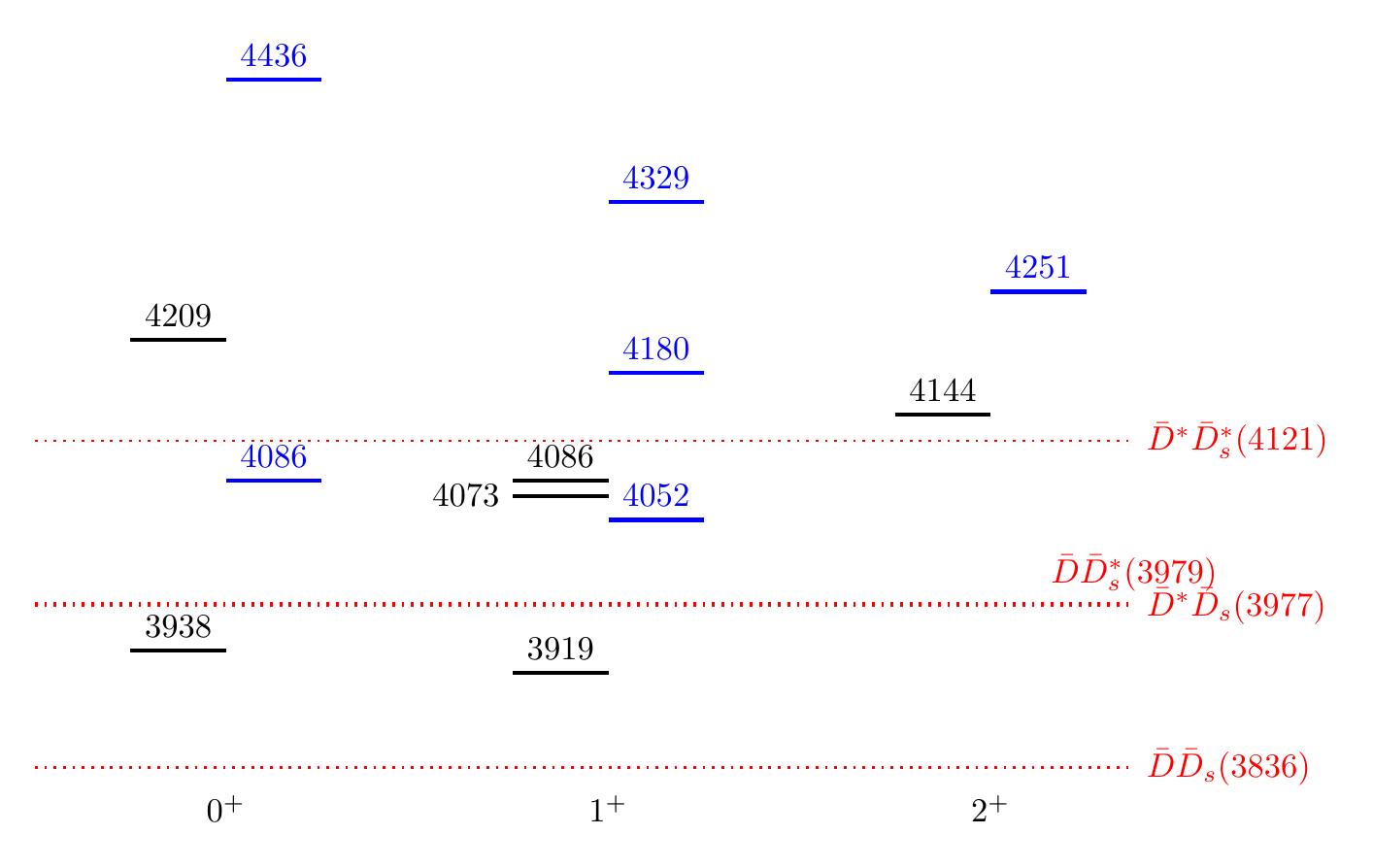}\\
		(a) $ns\bar{c}\bar{c}$ states\\
		&&\\
		\includegraphics[width=450pt]{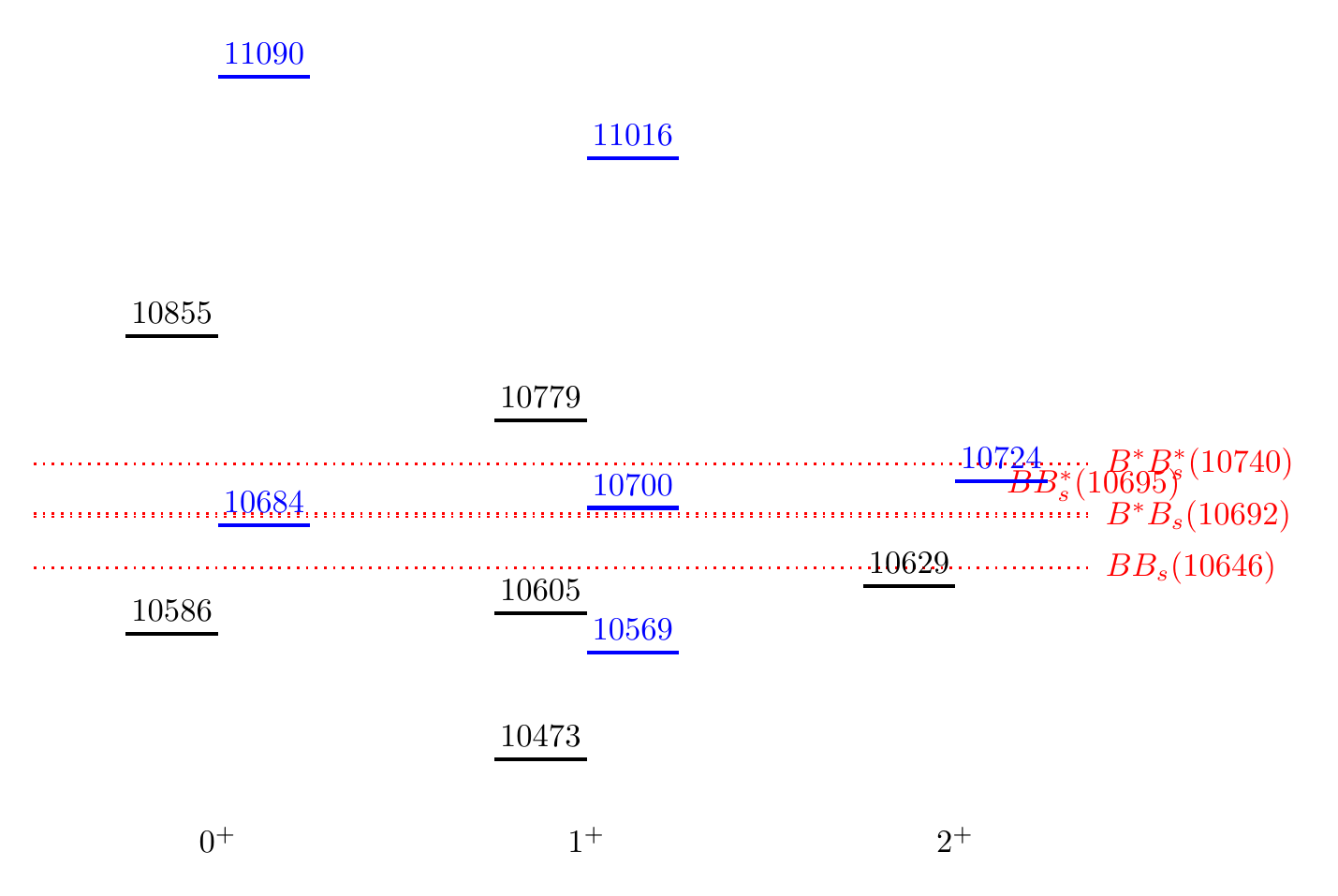}\\
		(b) $ns\bar{b}\bar{b}$ states\\
	\end{tabular}
	\caption{Mass spectra of the $ns\bar{c}\bar{c}$ and $ns\bar{b}\bar{b}$ tetraquark states in scheme~I (black) and scheme~II (blue). The dotted lines indicate various meson-meson thresholds. The masses are all in units of MeV.}
	\label{fig:nscc+nsbb}
\end{figure*}
%
%%%
%%% mass:nscb+:P1+P2
%%%
\begin{figure*}%[!h]
	\begin{tabular}{ccc}
		\includegraphics[width=450pt]{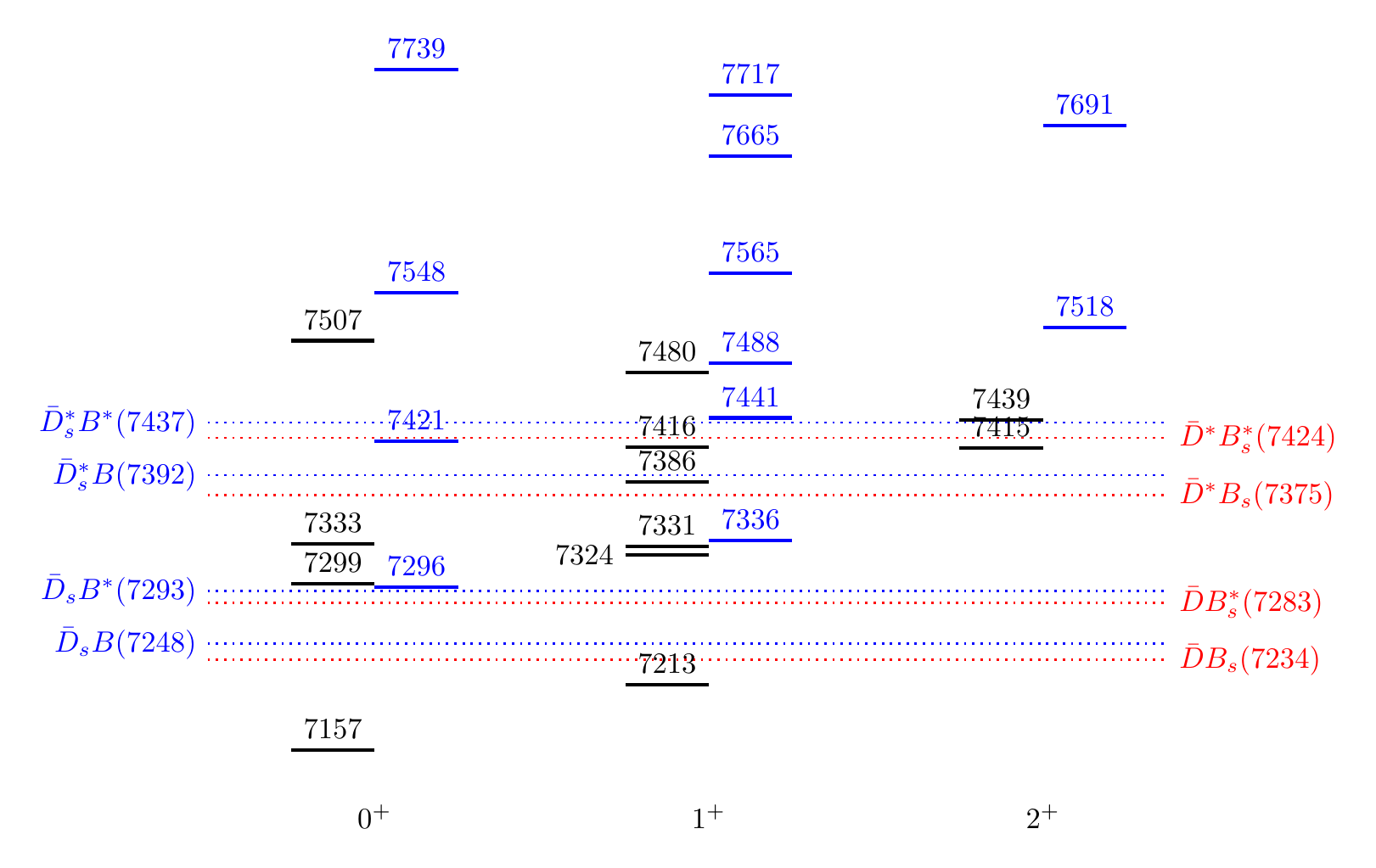}\\
	\end{tabular}
	\caption{Mass spectra of the $ns\bar{c}\bar{b}$ tetraquark states in scheme~I (black) and scheme~II (blue). The dotted lines indicate various meson-meson thresholds. The masses are all in units of MeV.}
	\label{fig:nscb}
\end{figure*}

We list the masses and wave functions of the $ns\bar{Q}\bar{Q}$ in Table~\ref{table:mass:nscc+nsbb+nscb:I+II}.
The ground states of the $ns\bar{c}\bar{c}$ and $ns\bar{b}\bar{b}$ tetraquarks are both of $1^{+}$.
They are strange counterparts of the $IJ^{P}=01^{+}$ $nn\bar{Q}\bar{Q}$ tetraquarks.
Among them, the $T_{I}(ns\bar{c}\bar{c},3919.0,1^{+})$ lies above the $\bar{D}\bar{D}_{s}$ threshold, while the $T_{I}(ns\bar{b}\bar{b},10473.1,1^{+})$ lies deeply below the $BB_{s}$ threshold.
In scheme~II, the former one lies above its $S$-wave decay channels $\bar{D}^{*}\bar{D}_{s}$ and $\bar{D}\bar{D}_{s}^{*}$, while the latter one is still stable.
We hope the future experiment can reach for this state.

The last class of the doubly heavy tetraquarks is the $ns\bar{c}\bar{b}$ system.
It is composed of four different quarks.
Similar to the $nn\bar{c}\bar{b}$ tetraquarks, the ground state of the $ns\bar{c}\bar{b}$ tetraquarks has quantum number $0^{+}$.
Depending on the scheme used, it may be a stable state.
A full dynamical quark model study is needed to have a better understanding of these states.

We also study the decay properties of the $ns\bar{Q}\bar{Q}$ tetraquarks, which can be found in Tables~\ref{table:eigenvector:nscc:I+II}--\ref{table:R:nscb:I+II}.

\newpage
%%%%%%%%%%%%%%%%%%%%%%%%%%%%%%%%%%%%%%%%%%%%%%%%%%%%%%%%%%%%%%%%%%%%
\section{Conclusions}
\label{Sec:Conclusion}
%%%%%%%%%%%%%%%%%%%%%%%%%%%%%%%%%%%%%%%%%%%%%%%%%%%%%%%%%%%%%%%%%%%%

In this work, we systematically study the mass spectrum of
the doubly heavy $qq\bar{Q}\bar{Q}$ tetraquarks in the framework of an extended
chromomagnetic model.
In addition to the chromomagnetic interaction, the effect of color
interaction is also considered in this model.
%
%The model parameters are determined from the mesons and baryons, which may differ from that of the tetraquarks.
%
The model parameters are fitted from the mesons and baryons.
Since the spatial configurations of the $qq$ ($\bar{q}\bar{q}$) and $q\bar{q}$ pairs are different in the conventional hadrons and the tetraquarks, applying these parameters to the tetraquarks may cause errors.
To appreciate this uncertainty, we adopt two schemes of parameters to study the $qq\bar{Q}\bar{Q}$ tetraquarks.
As indicated in Eq.~\eqref{eqn:HC:diff}, the scheme~II gives larger masses than the scheme~I.
However, the wave functions and decay properties of the two schemes are very similar for the $qq\bar{Q}\bar{Q}$ tetraquarks.
We find three states which are stable in both schemes.
They are the $nn\bar{b}\bar{b}$ tetraquark with quantum number $IJ^{P}=01^{+}$, the $nn\bar{c}\bar{b}$ tetraquark with quantum number $IJ^{P}=00^{+}$ and the $ns\bar{b}\bar{b}$ tetraquark with quantum number $J^{P}=1^{+}$.
They all lie below the thresholds of two pseudoscalar mesons, which can only decay through weak processes.
Meanwhile, many narrow states which lie below $S$-wave decay channels are also found.
It shall be interesting to search for these states.

The tetraquarks have two possible color configurations, namely the color-sextet configuration $\ket{(qq)^{6_{c}}(\bar{Q}\bar{Q})^{\bar{6}_{c}}}$ and the color-triplet one $\ket{(qq)^{\bar{3}_{c}}(\bar{Q}\bar{Q})^{3_{c}}}$.
Unlike the fully heavy tetraquarks, the ground states of the doubly heavy tetraquarks favor the color-triplet configurations.
Combining the results of fully and doubly heavy tetraquarks, we can clearly see the trend that the color-triplet configuration is more and more important when the mass ratio between the quarks and antiquarks increases.

Besides the mass spectrum, we also estimate the decay properties of the tetraquarks.
We hope these states can be searched for by future experiments.
%

%%%%%%%%%%%%%%%%%%%%%%%%%%%%%%%%%%%%%%%%%%%%%%%%%%%%%%%%%%%%%%%%%%%%
\section*{Acknowledgments}
%%%%%%%%%%%%%%%%%%%%%%%%%%%%%%%%%%%%%%%%%%%%%%%%%%%%%%%%%%%%%%%%%%%%

X.~Z.~W. is grateful to Marek~Karliner and Guang-Juan~Wang for helpful comments and discussions.
This project was supported by the National Natural Science Foundation of China (NSFC) under Grant No.~11975033 and No.~12070131001.
%

%%%%%%%%%%%%%%%%%%%%%%%%%%%%%%%%%%%%%%%%%%%%%%%%%%%%%%%%%%%%%%%%%%%%
\begin{appendix}
%%%%%%%%%%%%%%%%%%%%%%%%%%%%%%%%%%%%%%%%%%%%%%%%%%%%%%%%%%%%%%%%%%%%
\section{Wave function in the $q\bar{q}{\otimes}q\bar{q}$ configuration}
\label{app:wavefunc:13x24}
%%%%%%%%%%%%%%%%%%%%%%%%%%%%%%%%%%%%%%%%%%%%%%%%%%%%%%%%%%%%%%%%%%%%

To calculate the partial decay rates, we need to construct the tetraquark wave functions in the $q\bar{q}{\otimes}q\bar{q}$ configuration.
The possible color-spin wave functions $\{\beta_{i}^{J}\}$ are listed as follows,
\begin{enumerate}
\item $J^{P}=0^{+}$
\begin{align}
&\beta_{1}^{0}=\ket{\left(q_{1}\bar{q}_{3}\right)^{8}_{1}\left(q_{2}\bar{q}_{4}\right)^{8}_{1}}_{0}
\notag\\
%%%
&\beta_{2}^{0}=\ket{\left(q_{1}\bar{q}_{3}\right)^{8}_{0}\left(q_{2}\bar{q}_{4}\right)^{8}_{0}}_{0}
\notag\\
%%%
&\beta_{3}^{0}=\ket{\left(q_{1}\bar{q}_{3}\right)^{1}_{1}\left(q_{2}\bar{q}_{4}\right)^{1}_{1}}_{0}
\notag\\
%%%
&\beta_{3}^{0}=\ket{\left(q_{1}\bar{q}_{3}\right)^{1}_{0}\left(q_{2}\bar{q}_{4}\right)^{1}_{0}}_{0}
\end{align}
\item $J^{P}=1^{+}$
\begin{align}
	&\beta_{1}^{1}=\ket{\left(q_{1}\bar{q}_{3}\right)^{8}_{1}\left(q_{2}\bar{q}_{4}\right)^{8}_{1}}_{1}
	\notag\\
	%%%
	&\beta_{2}^{1}=\ket{\left(q_{1}\bar{q}_{3}\right)^{8}_{1}\left(q_{2}\bar{q}_{4}\right)^{8}_{0}}_{1}
	\notag\\
	%%%
	&\beta_{3}^{1}=\ket{\left(q_{1}\bar{q}_{3}\right)^{8}_{0}\left(q_{2}\bar{q}_{4}\right)^{8}_{1}}_{1}
	\notag\\
	%%%
	&\beta_{4}^{1}=\ket{\left(q_{1}\bar{q}_{3}\right)^{1}_{1}\left(q_{2}\bar{q}_{4}\right)^{1}_{1}}_{1}
	\notag\\
	%%%
	&\beta_{5}^{1}=\ket{\left(q_{1}\bar{q}_{3}\right)^{1}_{1}\left(q_{2}\bar{q}_{4}\right)^{1}_{0}}_{1}
	\notag\\
	%%%
	&\beta_{6}^{1}=\ket{\left(q_{1}\bar{q}_{3}\right)^{1}_{0}\left(q_{2}\bar{q}_{4}\right)^{1}_{1}}_{1}
\end{align}
\item $J^{P}=2^{+}$
\begin{align}
	&\beta_{1}^{2}=\ket{\left(q_{1}\bar{q}_{3}\right)^{8}_{1}\left(q_{2}\bar{q}_{4}\right)^{8}_{1}}_{2}
	\notag\\
	%%%
	&\beta_{2}^{2}=\ket{\left(q_{1}\bar{q}_{3}\right)^{1}_{1}\left(q_{2}\bar{q}_{4}\right)^{1}_{1}}_{2}
\end{align}
\end{enumerate}
where the superscript 1 or 8 denotes the color, and the subscript 0, 1 or 2 denotes the spin.
Among them, the $1_{c}{\otimes}1_{c}$ bases can also be written as combinations of two mesons.
For example, $\ket{\left(n_{1}\bar{c}_{3}\right)^{1}_{1}\left(n_{2}\bar{c}_{4}\right)^{1}_{1}}_{J}
{\equiv}\ket{\bar{D}^{*}\bar{D}^{*}}_{J}$.
%

%%%%%%%%%%%%%%%%%%%%%%%%%%%%%%%%%%%%%%%%%%%%%%%%%%%%%%%%%%%%%%%%%%%%
\end{appendix}
%%%%%%%%%%%%%%%%%%%%%%%%%%%%%%%%%%%%%%%%%%%%%%%%%%%%%%%%%%%%%%%%%%%%
\bibliography{myreference}
%\bibliography{G:/my_reference/myJabref}
%%%%%%%%%%%%%%%%%%%%%%%%%%%%%%%%%%%%%%%%%%%%%%%%%%%%%%%%%%%%%%%%%%%%
\end{document}